\pdfoutput=1
\documentclass[useAMS,usenatbib,twoside]{mn2e}
\usepackage[T1]{fontenc}
\usepackage{ae,aecompl}
\usepackage{times}
\usepackage{natbib}
\usepackage{lscape}
\usepackage{ulem}
\usepackage[usenames]{color}
\usepackage{graphicx}
\usepackage{xcolor} 
\usepackage{caption}
\usepackage{comment}
\usepackage{footnote}
\usepackage{amsmath}
\usepackage{siunitx}
\makeatletter
\let\jnl@style=\rm
\def\ref@jnl#1{{\jnl@style#1}}

\newcommand\ao{\ref@jnl{Appl.~Opt.}}
\newcommand\physrep{\ref@jnl{Phys.~Rep.}}
\newcommand\aj{\ref@jnl{AJ}}

\makeatother

\newcommand 	 \aap      {A\&A}

\newcommand 	 \apjs      {ApJS}
\newcommand 	 \apjl      {ApJL}

\newcommand       \msun        	{${\rm M}_{\odot}$}
\newcommand       \mic        	 {$\mu$m}
\newcommand       \lir          {L_{\rm IR}}
\newcommand       \applir        {\mu L_{\rm IR}}
\newcommand       \appd        {\sqrt{\mu}d}
\newcommand       \appMd     {\mu {M}_{d}}
\newcommand       \logten           {\log_{10}} 
\begin{document}

\title{On the redshift distribution and physical properties of ACT-selected DSFGs}
\author[T. Su et al.]{\Large T.~Su$^{1,}$\thanks{E-mail:ting@pha.jhu.edu}, T.A.~Marriage$^{1}$, V.~Asboth$^{2}$, A.J.~Baker$^3$, J.R.~Bond$^{4}$, D.~Crichton$^{1}$, M.J.~Devlin$^{5}$, R.~D{\"u}nner$^{6}$,
\newauthor \Large  D.~Farrah$^{7}$, D.T.~Frayer$^{8}$, M.B.~Gralla$^{1,9,10}$, K.~Hall$^{1}$, M.~Halpern$^{10}$, A.I.~Harris$^{11}$, M.~Hilton$^{12}$,
\newauthor \Large  A.D.~Hincks$^{2,13}$, J.P.~Hughes$^{3}$, M.D.~Niemack$^{14}$, L.A.~Page$^{15}$, B.~Partridge$^{16}$, J.~Rivera$^3$, D.~Scott$^{2}$, 
\newauthor \Large J.L.~Sievers$^{17}$, R.J.~Thornton$^{18}$, M.P.~Viero$^{19}$, L. Wang$^{20}$, E.J.~Wollack$^{21}$, M. Zemcov$^{22,23}$
\\
\footnotesize $^1$ Department of Physics and Astronomy, The Johns Hopkins University, 3400 N. Charles St., Baltimore, MD 21218-2686, USA\\
$^2$ Department of Physics and Astronomy, University of British Columbia, 6224 Agricultural Rd., Vancouver BC V6T 1Z1, Canada\\
$^3$ Department of Physics and Astronomy, Rutgers, The State University of New Jersey, 136 Frelinghuysen Road, Piscataway, NJ 08854-8019, USA\\
$^4$ Canadian Institute for Theoretical Astrophysics, University of Toronto, Toronto, ON, M5S 3H8, Canada \\
$^5$ Department of Physics and Astronomy, University of Pennsylvania, 209 South 33rd Street, Philadelphia, PA 19104 USA \\
$^6$ Departamento de Astronom{\'i}a y Astrof{\'i}sica, Pontific{\'i}a Universidad Cat{\'o}lica, Casilla 306, Santiago 22, Chile\\
$^7$ Department of Physics, Virginia Tech, Blacksburg, VA  24061, USA\\
$^8$ National Radio Astronomy Observatory, P.O. Box 2, Green Bank, WV 24944, USA \\
$^9$ Harvard-Smithsonian Center for Astrophysics, 60 Garden Street, Cambridge, MA 02138, USA\\
$^{10}$ Department of Astronomy/Steward Observatory, University of Arizona, 933 North Cherry Avenue, Tucson, AZ 85721-0065, USA \\
$^{11}$ Department of Astronomy, University of Maryland, College Park, MD 20742-2421, USA\\
$^{12}$ Astrophysics and Cosmology Research Unit, School of Mathematics, Statistics and Computer Science,
University of KwaZulu-Natal, Durban 4041, \\  South Africa\\
$^{13}$ Department of Physics, University of Rome `La Sapienza', Piazzale Aldo Moro 5, I-00185 Rome, Italy \\
$^{14}$ Department of Physics, Cornell University, Ithaca, NY 14853, USA\\
$^{15}$ Joseph Henry Laboratories of Physics, Jadwin Hall, Princeton University, Princeton, NJ 08544, USA\\
$^{16}$ Department of Astronomy, Haverford College, Haverford, PA 19041, USA\\
$^{17}$ Astrophysics and Cosmology Research Unit, School of Chemistry and Physics, University of KwaZulu-Natal, Durban 4041, South Africa\\
$^{18}$ Department of Physics, West Chester University, 700 S High St, West Chester, PA 19382, USA\\
$^{19}$ Kavli Institute for Particle Astrophysics and Cosmology, Stanford University, 382 Via Pueblo Mall, Stanford, CA 94305, USA\\
$^{20}$ SRON Netherlands Institute for Space Research, Landleven 12, 9747 AD, Groningen, The Netherlands\\
$^{21}$ NASA/Goddard Space Flight Center, Greenbelt, MD, 20771, USA\\
$^{22}$ Center for Detectors, School of Physics and Astronomy, Rochester Institute of Technology, 1 Lomb Memorial Dr., Rochester NY 14623, USA\\
$^{23}$ Jet Propulsion Laboratory, 4800 Oak Grove Dr., Pasadena CA 91109, USA}
\maketitle

\begin{abstract}
We present multi-wavelength detections of nine candidate gravitationally-lensed dusty star-forming galaxies (DSFGs) selected at 218\,GHz (1.4\,mm) from the ACT equatorial survey. Among the brightest ACT sources, these represent the subset of the total ACT sample lying in  \textit{Herschel} SPIRE fields, and all nine of the 218\,GHz detections were found to have bright \textit{Herschel} counterparts. By fitting their spectral energy distributions (SEDs)  with a modified blackbody model with power-law temperature distribution, we find the sample has a median redshift of $z=4.1^{+1.1}_{-1.0}$ (68 per cent confidence interval), as expected for 218\,GHz selection, and an apparent total infrared luminosity of $\log_{10}(\applir/{\rm L}_\odot) = 13.86^{+0.33}_{-0.30}$, which suggests that they are either strongly lensed sources or unresolved collections of  unlensed DSFGs. The effective apparent diameter of the sample is $\appd = 4.2^{+1.7}_{-1.0}$\,kpc, further evidence of strong lensing or multiplicity, since the typical diameter of dusty star-forming galaxies is $1.0$--$2.5$ kpc. We emphasize that the effective apparent diameter derives from SED modelling without the assumption of optically thin dust (as opposed to image morphology). We find that the sources have substantial optical depth ($\tau = 4.2^{+3.7}_{-1.9}$) to dust around the peak in the modified blackbody spectrum ($\lambda_{\rm obs} \le 500$\,$\mu$m), a result that is robust to model choice.

\end{abstract}
\begin{keywords} 
galaxies: evolution -- galaxies: formation -- galaxies: high-redshift -- galaxies: starburst -- submillimetre: galaxies
\end{keywords}

\section{Introduction}
\label{sec:intro}

Observations of the light from young massive stars, whether directly in the rest-frame UV or after reprocessing by dust in the far-infrared/submillimetre, have allowed us to map out the cosmic history of star formation \citep[e.g.,][]{Lilly96,Madau96,Blain02,Chapman05,LeFloch05,Perez-Gonzalez05,Hopkins06,Daddi07,Elbaz07,Casey14,Madau14}. Due to these observations we now know that the Universe formed most of its stars in the redshift range $1<z<3$. However, our understanding of cosmic star-formation during this epoch is still incomplete. As a tracer of star formation rate (SFR), the rest-frame UV is reprocessed by dust in intermediate/high redshift galaxies as the more intense star-forming activity in these early epochs is often associated with dusty environments, especially in the most extreme star-forming systems.
Therefore, infrared/submillimetre re-emission by dust from the star-forming regions is crucial when accounting for SFRs at intermediate/high redshift. Furthermore, a large number of observations indicate that the so-called classical submillimetre galaxies (SMGs) play a key role in galaxy evolution as the likely progenitors of today's massive elliptical galaxies \citep{Blain02,Casey14}.
Our understanding of the physical properties of SMGs is insufficient due to the lack of large enough samples with detailed multi-wavelength data. The samples and associated data, however, are rapidly improving. In particular, a new population of lensed dusty star-forming galaxies (DSFGs) selected in large ($>100$~square degrees) millimetre and submillimetre surveys has begun to enable  a closer look at star formation in the high-redshift universe.

Over the past five years, new samples of gravitationally lensed DSFGs have been detected at millimetre and submillimetre wavelengths with the Atacama Cosmology Telescope \citep[ACT;][]{marsden14}, {\it Herschel} \citep[e.g.][]{Negrello10,Conley11,Wardlow13}, {\it Planck} \citep{Canameras15,Harrington16}, and the South Pole Telescope \citep[SPT;][]{Vieira10,Mocanu13}. 
Unlike with optical selection, which relies on morphological identification of lensed galaxies near the Einstein radius (through sheared or multiple images), the selection of these lensed DSFGs is based only on apparent flux (through magnification). This selection is possible due to the steep decline in the number density of DSFGs with flux. The brightest observed sources stand out because of lensing or unresolved mergers where many bright unlensed DSFGs  are confused for a single ultra-bright source \citep[i.e., ``trainwrecks'';][]{Riechers11,Ivison12,Fu13}. 
Because of the rarity of lensed galaxies, a blind millimetre/submillimetre search with requisite sensitivity ($1$--$10$ mJy flux limits) over a large survey area is required, and this is exactly what has been achieved by ACT, {\it Herschel}, {\it Planck}, and SPT. Besides lensed DSFGs, blazars are bright at millimetre wavelengths, but millimetre-wave spectral indices and information from longer-wavelength radio surveys can be used to veto these systems. Interferometric follow-up observations with the Submillimeter Array (SMA), the Atacama Large Millimeter/submillimeter Array (ALMA), and other observatories confirm that the majority of these DSFGs are strongly lensed  \citep[e.g.,][]{Negrello10,Bussmann12,Vieira13,Hezaveh13,Spilker16}.

In addition to providing an efficient means of finding lensed systems at millimetre and submillimetre wavelengths, the magnification allows us to probe the internal structure of high-$z$ DSFGs. When the light from a distant starburst galaxy is lensed by a foreground object, the apparent effective size and the apparent luminosity of the background source are magnified by factors of $\sqrt{\mu}$ and $\mu$, respectively. For unlensed sources $\mu=1$, while for strongly lensed sources $\mu\ge2$. For instance, the "Cosmic Eyelash", which is a lensed DSFG at $z=2.3259$ \citep{Swinbank10}, can be resolved at a scale of 100 pc owing to an extreme magnification $\mu \approx 32$ by a foreground cluster lens. In another study, the morphology of a highly magnified ($\mu \approx 21$) source, SPT 0538$-$50 at $z=2.7817$ indicates a merger-driven event, which is similar to a local ultraluminous infrared galaxies (ULIRGs) \citep{Bothwell13}. In a third, \citet{Fu12} present a detailed study of HATLAS~J114637.9$-$001132 at $z=3.2592$ with multi-wavelength images revealing different magnification factors for stars, dust and gas. Additionally, \citet{Hezaveh13} and \cite{Spilker16} study strongly lensed DSFGs at $z=1.9-5.6$, recovering accurate intrinsic source sizes and source surface brightness densities on scales not achievable in the absence of lensing. In another detailed study, HATLAS~J090311.6+003906 \citep[SDP.81;][]{Negrello10} at $z = 3.042$  ($\mu \approx 11$) has been observed at a resolution of 23\,mas, corresponding to a $\sim180$\,pc physical scale. It has molecular and dust clumps confined to a $\sim 2$\,kpc region, while its stellar component occupies a larger volume offset from the dust \citep{Bussmann13,Dye14,ALMAPartnership15, Dye15, Rybak15a,Rybak15b, Wong15, Tamura15, Hatsukade15, Rybak15b}. Resolved spectroscopy suggests that SDP.81 is undergoing a merger-driven starburst phase \citep{Hatsukade15, Rybak15a, Dye15}.

While lensed DSFGs are discovered in surveys with modest resolution ($0.3$--$5'$) and large areas, the detailed study of lensed DSFGs involves a variety of challenging observations, including high resolution millimetre/submillimetre imaging, detection of multiple CO lines for spectroscopic redshift determination, and optical or near-infrared studies of the lenses (both imaging and spectroscopy). Even without these follow-up data, however, the far-infrared (FIR) spectral energy distribution (SED) of the thermal dust emission, obtained through photometry of available survey data, allows us to estimate dust mass, dust temperature, infrared luminosity and SFR, given knowledge of the redshift \citep[e.g.,][]{Negrello10,Cox11, Bussmann13,Bothwell13,Wardlow13}. These studies generally use either template SEDs derived from fiducial starburst galaxies or single-temperature modified blackbody SED models with or without the assumption that the dust is optically thin over the range of frequencies probed. Conversely, if one has constraints on the dust mass and temperature, then a reasonably constrained photometric redshift can be obtained. For instance, \citet{Greve12} estimate the redshift distribution of SPT-selected sources by fitting their SED data using a modified blackbody model with a fixed single temperature. 

\begin{table*}
 \begin{center}
  \begin{tabular}{@{}ccccccccc@{}}
  %\hline
     \multicolumn{1}{c}{}   &\multicolumn{2}{c}{}&\multicolumn{3}{|c|}{ACT}& \multicolumn{3}{c|}{\textit{Herschel} SPIRE}\\

 ACT ID & RA &Dec& 148\,GHz & 218\,GHz & 278\,GHz & 500\,$\mu$m &350\,$\mu$m& 250\,$\mu$m\\
	    & [deg]  & [deg]      &	[mJy]	       &[mJy]		 &[mJy]			&[mJy]		&[mJy]			&[mJy]		\\
  \hline
ACT-S J0011$-$0018  &2.8902 &  -0.3101 & $6.02\pm  2.23$  &$ 21.93\pm3.11$  &$28.43\pm9.74$     & $94.65\pm7.41$ &$120.18\pm6.33$ & $89.41\pm 7.25$ \\
ACT-S J0022$-$0155  &5.5870 & -1.9230  & $5.54 \pm2.58$  &$ 26.02\pm4.08$  & ---                & $116.35\pm6.57$&$104.05\pm5.84$   &$62.67\pm6.10$\\
ACT-S J0038$-$0022  &9.5586 & -0.3810	& $7.95\pm1.86 $  &$23.91\pm  2.65$  &$ 40.88\pm 5.88 $  & $122.87\pm6.69$&$119.01\pm6.01$   &$73.35\pm5.55$\\
ACT-S J0039+0024    &9.8723 & 0.40736  & $6.14\pm 1.76$  &$ 19.50\pm2.56$  &$35.32\pm 6.24$     & $162.11\pm 7.29 $&$152.60\pm6.34$ &$140.84\pm6.51$\\
ACT-S J0044$+$0118  &11.0421 & 1.3071	 & $12.49\pm 1.74$ &$ 35.11\pm  2.62$  &$ 72.32 \pm 6.26$ & $191.79\pm8.16$&$166.48\pm5.99$ &$108.14\pm6.86$\\
ACT-S J0045$-$0001  &11.3860 &-0.0232 & $ 6.46 \pm 1.90$  &$ 24.99 \pm  2.51$  &$45.87\pm6.23$ & $97.18\pm7.36$ &$87.61\pm6.28$  &$48.22\pm6.73$\\
ACT-S J0107+0001    &16.8709 &2.2439  & $1.77\pm 2.00 $   &$ 18.26\pm2.74$  &$ 28.72\pm 6.08$  &$87.17\pm7.70$ &$ 89.76\pm6.44$ &$48.55\pm6.11$\\
ACT-S J0116$-$0004  &19.1670 &-8.1598 & $ 7.69\pm1.91  $    &$ 22.73\pm2.60$ &$45.99 \pm5.82 $  &$189.91\pm7.81  $ &$196.11\pm6.56 $ &$ 137.50\pm6.75$\\
ACT-S J0210+0016    &32.4215 &2.6599 & $17.06\pm1.64  $   &$   69.25\pm2.68     $ &$154.15 \pm5.67 $ & $ 717.64\pm8.13 $ &$ 912.36\pm6.57$ &$ 826.02\pm6.68 $\\
\hline
\end{tabular}
\end{center}

 \caption{Coordinates and photometric data for ACT-selected lensed DSFG candidates. The coordinates are extracted from {\it Herschel} SPIRE 250\,$\mu$m photometry. Three sources (ACT-S J0107+0001, ACT-S J0116$-$0004 and ACT-S J0210+0016) have {\it Herschel} SPIRE data from HerS \citep{Viero14}. The other six have {\it Herschel} SPIRE data from HeLMS \citep{Oliver12,Asboth16}. Errors in the table are only due to instrument noise. Additional uncertainties due to systematic effects and confusion are enumerated in Section \ref{sec:data} and are included in the modelling. The footprint of the 278\,GHz data does not cover the coordinates of ACT-S J0022$-$0155.}
\label{ACTtable}
\end{table*}

This paper represents a first look at the physical properties and redshift distribution for DSFGs selected by ACT. This is one of two growing samples of DSFGs selected in millimetre-wave surveys over thousands of square degrees of sky. (The other sample is from SPT.) We derive physical properties through SED modelling of nine candidate gravitationally lensed DSFGs selected from the ACT equatorial survey (Gralla et al., in prep) in the overlap region with the Herschel Stripe 82 Survey \citep[HerS;][]{Viero14} and the HerMES Large Mode Survey \citep[HeLMS;][]{Oliver12}. We explore a variety of SED models, but we focus on a fiducial modified blackbody model without the assumption of optically thin dust and with a power-law dust temperature distribution. Such a model is well suited for characterizing the SEDs of high-$z$ DSFGs \citep[e.g.][]{Blain03,Kovacs06,Kovacs10,Magnelli12,Casey12,Bianchi13,DaCunha13,Staguhn14}. The power-law temperature distribution better captures non-trivial spectral properties at the peak and on the Wien side of the modified blackbody that are not described well by single-temperature models \citep{Kovacs10,Magnelli12}. In particular, the power-law temperature distribution better captures the rest frame mid-infrared emission coming from the smaller clumps of hotter dust near the galaxy nucleus \citep{Kovacs10,Casey14}. By comparing the goodness of fit of our fiducial model with those of optically thin models, we will test whether the sources are optically thin at all observed wavelengths.

The paper is organized as follows. In Section \ref{sec:data}, we describe the ACT DSFG sample selection, the photometric data used to construct the SEDs, and auxilliary data. In Section \ref{sec:modeling}, we explain the SED model and fitting methods. The results and discussion are given in Section \ref{sec:results}. In Section \ref{sec:conclusions}, we lay out the conclusions. We adopt a flat $\Lambda$CDM cosmology, with a total matter (dark+baryonic) density parameter of 
$\Omega_{\rm m}=0.27$, a vacuum energy density $\Omega_{\Lambda}=0.73$, and a Hubble constant of $H_0 = 70~$\,km\,s$^{-1}$\,Mpc$^{-1}$.

\section{Observations and data}
\label{sec:data}

\subsection {The Atacama Cosmology Telescope}
\label{sec:act}

ACT is a 6~m telescope in the Atacama Desert that operates at millimetre wavelengths \citep{Fowler07,Swetz11,niemack10,Thornton16}. The ACT data for this study were collected in 2009 and 2010 at 148\,GHz (2\,mm), 218\,GHz (1.4\,mm) and 277\,GHz (1.1\,mm). They cover 480 $\deg^2$ on the celestial equator with right ascension  $-58^\circ < \alpha < 57^\circ$ and declination $-2.2^\circ<\delta<2.2^\circ$. Intensity maps at the three frequencies were made as described in \citet{dunner13}, with resolutions of  1.4$'$ (148\,GHz), 1.0$'$ (218\,GHz), and 0.9$'$ (227\,GHz), respectively \citep{hasselfield13}. The maps were match filtered and sources identified as in \citet{marsden14}. The resulting flux densities have typical statistical errors due primarily to instrument noise of 2.2\,mJy, 3.3\,mJy, and 6.5\,mJy and fractional systematic errors due to calibration, beam errors, frequency-band errors, and map-making  at the  3 per cent, 5 per cent, and 15 per cent level for 148\,GHz, 218\,GHz and 277\,GHz, respectively \citep{Gralla14}.

\begin{figure*}
  \centering
 \includegraphics[width=7.0in]{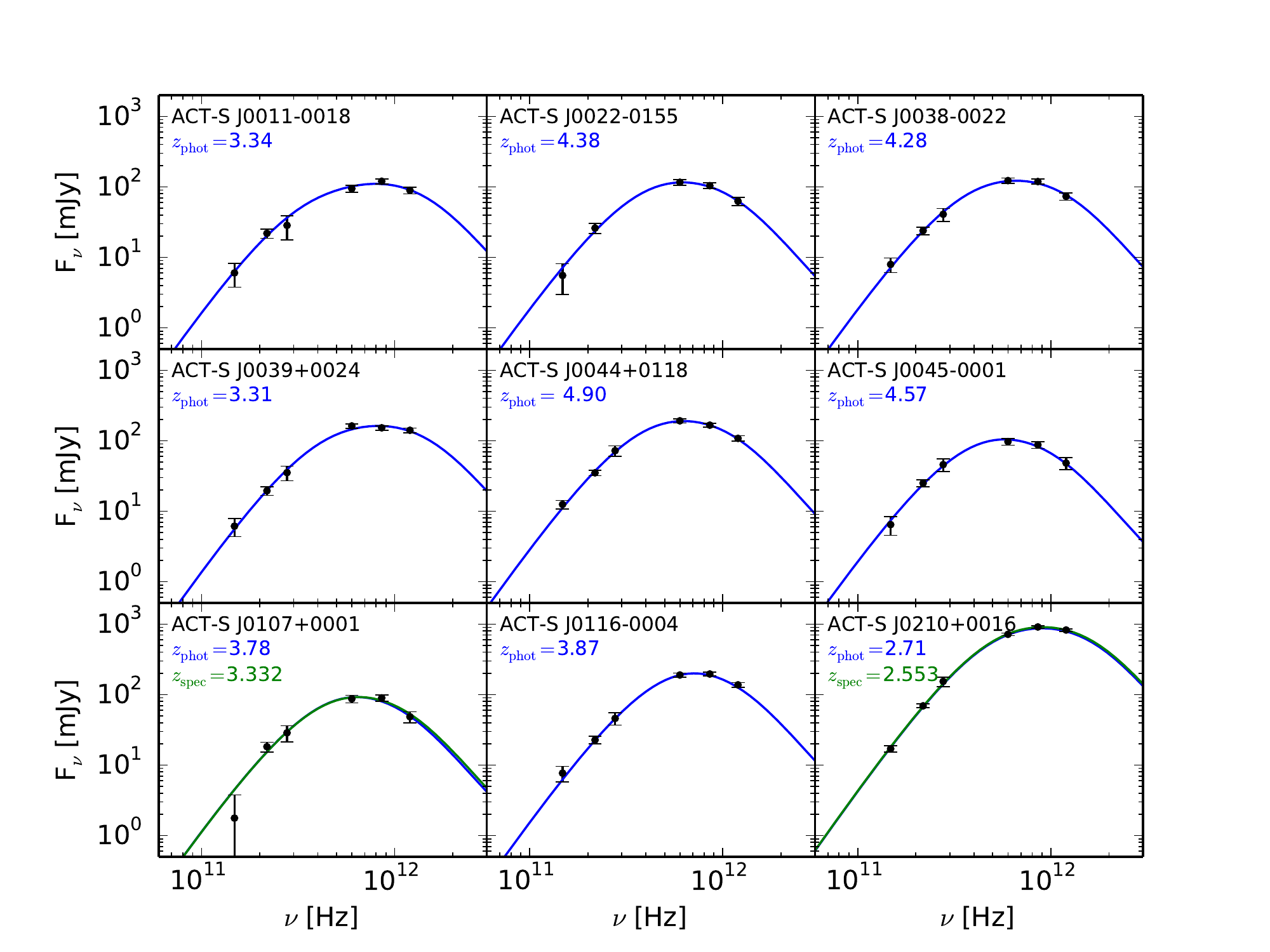} 
 \caption{SED data and best fit models for the ACT-selected lensed DSFG candidates. As plotted, the errors include  systematic error (e.g., calibration uncertainties) in addition to instrument and confusion noise.  The blue curves are the best-fitting results of the power-law temperature modified blackbody model without the assumption of optically thin emission (our fiducial model of Equation \ref{multi_model}). For sources ACT-S J0107+0001 and ACT-S J0210+0016, the green curve shows the results of our fit if the redshift parameter is fixed to the measured spectroscopic redshift. Additional models are explored in Section~\ref{sec:other_models} and Appendix~\ref{SEDfitting}.}
 \label{SEDs}
\end{figure*}

From  the filtered 218\,GHz data, we selected the thirty brightest 218\,GHz DSFG candidates for further study and follow-up observations.  The DSFGs were selected to have a 148-218~GHz spectral index $(S\propto \nu^\alpha)$ consistent with that of thermal dust $(\alpha>2)$ and inconsistent with blazar spectra. Candidates were vetoed if they had counterparts corresponding to nearby star-forming galaxies resolved in optical imaging from the Sloan Digital Sky Survey (SDSS). Candidates were also rejected if they fell in regions of the map contaminated by Galactic cirrus. The full description of the sample selection will be given in Gralla et al. (in prep). Of these thirty brightest DSFG candidates selected in ACT data, nine fall in a 120 deg$^2$ region of the sky also observed by {\it Herschel}, as described in Section \ref{sec:herschel}. The ACT flux densities for these nine sources are given in Table \ref{ACTtable} and plotted in Figure~\ref{SEDs}. Note that the overlap with {\it Herschel} data is centered on a deep part of the ACT map, and thus the statistical flux density errors given in Table \ref{ACTtable} are, on average, below the typical error quoted in the previous paragraph for the data as a whole. These flux densities are the raw values and thus have not been corrected for Eddington bias. However, given that all candidates have 218\,GHz flux densities in excess of 18\,mJy and signal-to-noise S/N~$>$~6, this correction will be less than a few per cent \citep{marsden14} and will not affect the results of our SED analysis at a significant level. Because the sources are selected at 218~GHz with  instrument noise dominant and uncorrelated between bands, the level of Eddington bias is primarily determined by the 218~GHz selection, and the lower S/N of the other ACT bands does not lead to more Eddington bias. In addition to Eddington bias, the flux may be boosted due to selecting peaks in S/N in the match filtered map \cite[e.g.,][]{Vanderlinde10}. Specifically, by optimizing the S/N over the right ascension and declination coordinates, we bias the detected source flux  by a factor \mbox{(S/N)/((S/N)$^2$-2)$^{1/2}$}. Correcting for this bias changes the best-fit values of our SED model parameters by much less than the corresponding model errors. Therefore, for simplicity, the model results presented in Section \ref{sec:results} are based on fits to the raw flux densities given in Table \ref{ACTtable}.

To estimate the astrometry of the ACT DSFG sample, we compare the ACT-derived source locations to those obtained through high resolution SMA follow-up of the sources. (See Section \ref{sec:data_other}.) The ACT-derived astrometric uncertainty thus derived is $6-7''$. 

\subsection {The {\it Herschel Space Observatory}}
\label{sec:herschel}

Nine ACT-selected DSFG candidates fall within the region where the ACT survey overlaps  HerS and HeLMS, an area of 120 deg$^2$ in the right ascension range $0^\circ<\alpha<37^\circ$. The corresponding submillimetre data from the {\it Herschel} Spectral and Photometric Imaging REceiver \citep[SPIRE;][]{griffin10} can help constrain the turn-over of the SED of thermal dust in these galaxies if they have redshifts $z>1.5$ and dust temperatures $T>20$~K.
SPIRE observes at wavelengths of 250\,$\mu$m (1200\,GHz), 350\,$\mu$m (857\,GHz) and 500\,$\mu$m (600\,GHz) with corresponding beam sizes of 0.3$'$, 0.4$'$ and 0.6$'$.  The SPIRE flux calibration uncertainty of 4 per cent and beam full width at half maximum (FWHM) values are derived from observations of Neptune \citep{griffin13}. 

For each of the nine ACT-selected sources with a potential {\it Herschel} counterpart we consulted publicly available catalogs from HerS \citep[][]{Viero14}\footnote{www.astro.caltech.edu/hers/HerS\_Home.html} and HeLMS \citep[][]{Asboth16}\footnote{http://hedam.lam.fr/HerMES/}. (See also \cite{Nayyeri16}.) Here we summarize the construction of each of these catalogs, and we refer the reader to the corresponding papers for complete information on catalog construction. Point-source extraction was performed on HerS maps after first filtering them with a tapered high-pass filter to remove large-scale Galactic cirrus. Sources were identified in the $250$~$\mu$m image using the IDL software package StarFinder \citep{Diolaiti00}, and photometry extracted from all three bands using a modified version of the DESPHOT algorithm \citep{Roseboom10,Roseboom12,Wang14}. The benefit of this approach is that it uses input sources from the highest resolution band as a prior for the other SPIRE wavelengths,  thus producing consistent, band-merged SPIRE catalogues. The HeLMS flux densities and  uncertainties used in this work come from the HeLMS red source catalogue \citep{Asboth16}. Sources were extracted from a linear combination of the  500~$\mu$m HeLMS map, match filtered as in \cite{Chapin11},  and the 250~$\mu$m HeLMS map, smoothed to the resolution of the filtered 500~$\mu$m data. The linear combination was chosen to minimize variance due to the common confusion noise in both maps as in \cite{Dowell14}. The flux densities for the sources thus detected are found by performing an inverse-variance-weighted convolution of each of the SPIRE maps with the point spread function, similar to the method described in \cite{Smith11}.

The errors in \textit{Herschel} SPIRE flux densities derive from instrument noise, confusion noise, and the aforementioned calibration uncertainty. We sum these contributions in quadrature to obtain the total photometric uncertainties used in the modelling. The characteristic rms instrument noise level in the {\it Herschel} data is roughly 6\,mJy for most measurements, as seen in Table \ref{ACTtable}. The confusion noise is 6.8\,mJy (500\,$\mu$m), 6.2\,mJy (350\,$\mu$m) and 5.8 (250\,$\mu$m) \citep{Nguyen10}. Each of our nine ACT sources in the \textit{Herschel} survey areas has significant flux density in all {\it Herschel} SPIRE bands. 

In addition to submillimetre flux densities, one can obtain accurate 250~$\mu$m-based positions for the sources from the {\it Herschel} SPIRE catalogs \citep{Viero14,Asboth16}. The astrometric uncertainty for the {\it Herschel} SPIRE positions as referenced to our SMA follow-up is approximately $3''$. Comparing these positions to the ACT-derived astrometry, we find differences in right ascension (declination) with mean and standard deviation of $-0.8''\pm3.2''$ ($-1.23''\pm2.33''$). The two datasets are consistent in terms of astrometry, and no ambiguity exists in terms of source cross-identification. Given the better astrometry of the Herschel data, we use these locations in Table \ref{ACTtable}.

\subsection {Additional data}
\label{sec:data_other}

The ACT sample of candidate lensed DSFGs has been the subject of a campaign of multi-wavelength follow-up observations, from the radio to the optical. In this study we primarily restrict our attention to the integrated flux density data from ACT and {\it Herschel}. One exception is that we compare our results to spectroscopic redshift measurements for ACT-S J0210$-$0016 and ACT-S J0107+0001. For ACT-S J0210$-$0016 we use a spectroscopic (CO-based) redshift of $z=2.553$ from  our follow-up observations with the Green Bank Telescope (GBT) and the Combined Array for Research in Millimeter-wave Astronomy (CARMA), which we present in Appendix \ref{app:0210}. For ACT-S J0107+0001, we used a CO-based spectroscopic redshift of $z = 3.332$ from a measurement with the Redshift Search Receiver \citep[RSR;][]{Erickson07} on the Large Millimeter Telescope (LMT); a paper discussing this and other LMT/RSR observations of ACT DSFGs is currently in preparation.

\subsubsection{Beyond FIR photometry}

The FIR photometry analyzed in this study provides certain insights into the ACT-selected sample, to be discussed in the remainder of this paper. Much more information, however, can be gained with additional archival data and follow-up observations. Therefore, as a supplement to the present study and as a prelude to future work, we present optical, near-IR, mid-IR, and high-resolution submillimetre data on our nine sources.

For 218~GHz-selected lensed source populations, the associated lenses are expected to be massive elliptical galaxies and galaxy clusters with a broad redshift distribution extending up to $z\approx1.5$ \citep{Hezaveh11}. Over this broad redshift range, a complete lens assay requires a full complement of optical, near-IR and mid-IR  observations. We use optical imaging data from the SDSS \citep{York00,Gunn06,Eisenstein11,Alam15}. In the near-IR we use data from two sources: the VISTA Hemisphere Survey \cite[VHS;][]{McMahon13} with a Ks-band ($\lambda \sim 2.1$~$\mu$m) 5-sigma detection limit of $m=18.1$ (Vega) and our own follow-up observations with the NICFPS camera on the ARC 3.5~m telescope at the Apache Point Observatory \citep{Vincent03} with a Ks-band 5-sigma detection limit of $m=19.5$ (Vega). We also use imaging at 3.6~$\mu$m and 4.5~$\mu$m from the  {\it Spitzer} IRAC Equatorial Survey \citep{Timlin16} and the {\it Spitzer}-HETDEX Exploratory Large-Area Survey \citep{Papovich16}. Where {\it Spitzer} data is unavailable, we use mid-IR data from the  {\it Wide-field Infrared Survey Explorer} ({\it WISE}) all-sky survey \citep{Wright10}. 

\begin{figure*}
  \centering
 \includegraphics[width=5.5in]{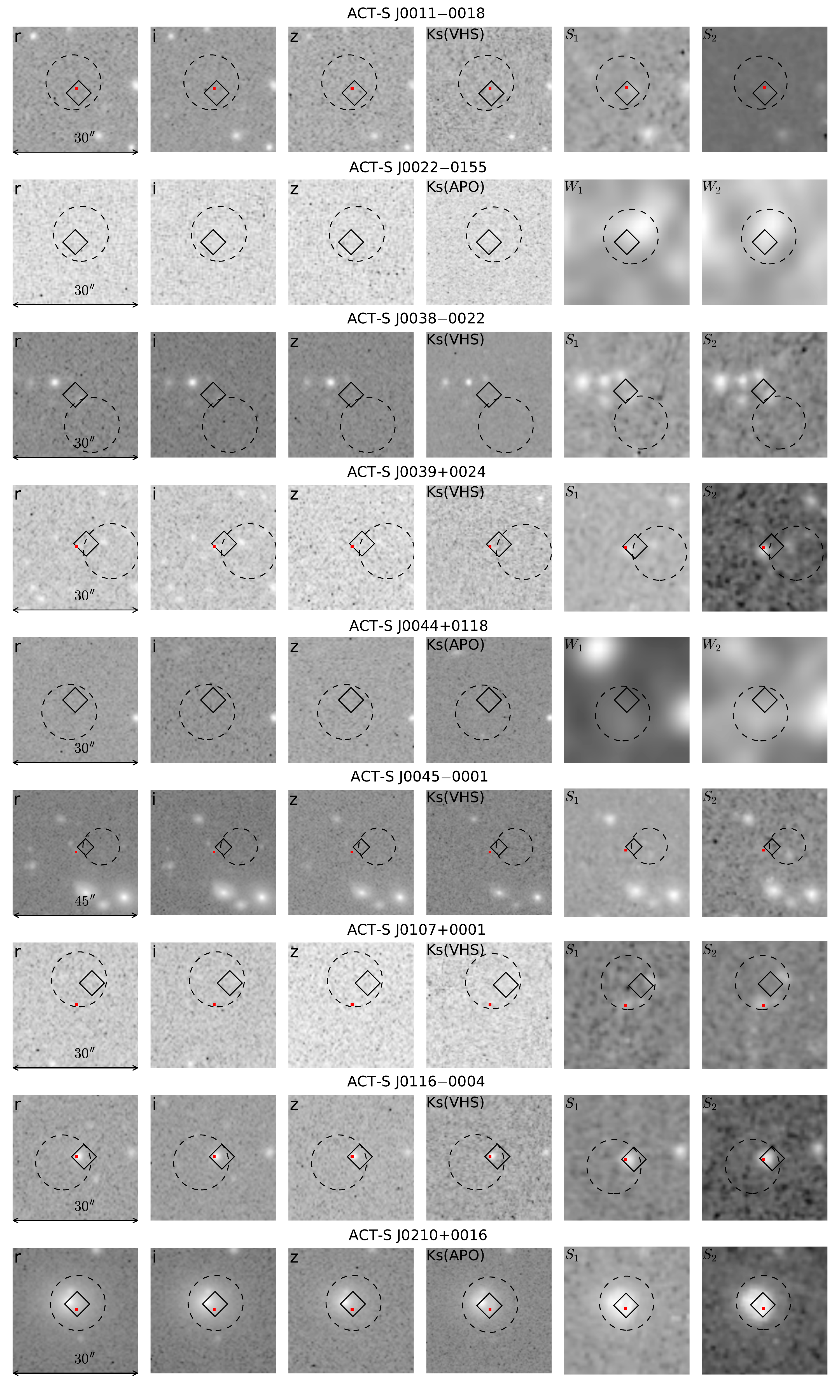} 
 \caption{Gray-scale images are optical (SDSS), near-IR (VHS or APO/ARC~3.5~m), and mid-IR ({\it Spitzer} or {\it WISE}). The most accurate source location (red square) comes from our high-resolution submillimetre follow-up imaging with the SMA, where available. The {\it Herschel} and ACT-derived locations are shown, respectively, by a solid diamond and dashed circle, the sizes of which indicate astrometric error.}
 \label{fig:additional_data}
\end{figure*}

We have an on-going program with the Submillimeter Array (SMA; Programs 2013B-S066, 2015B-S049) to image ACT-selected DSFGs with 3$''$ resolution at 230~GHz (Rivera et al. in prep). These follow-up observations provide improved astrometry for optical/IR counterpart identification. The data  also distinguish between lensing and ``trainwreck'' merger scenarios. In this work, we make use of the improved astrometry for six of the nine sources to assess the relationship of the ACT-selected sources with galaxies detected in the optical/IR imaging. Figure \ref{fig:additional_data} shows the result. For most sources, the location of the ACT-selected source is consistent with lensing either by a galaxy or galaxy cluster detected in the IR/optical data. 

For putative lens galaxies or galaxy clusters, we use SDSS to establish spectroscopic and photometric redshifts. For lens candidates lacking redshifts from SDSS, we have an on-going spectroscopic follow-up campaign with the South African Large Telescope \citep[SALT;][]{Buckley06}  to determine  redshifts (PI J. Hughes). Below we give a source-by-source description of the optical/IR data in Figure~\ref{fig:additional_data} with lens candidate redshift estimates where available.

\begin{itemize}
\item \mbox{ACT-S J0011$-$0018}  has an accurate SMA location allowing identification of a  nearby lens candidate apparent only in the mid-IR data. 

\item \mbox{ACT-S J0022$-$0155} does not have SMA data, but  \textit{Herschel} astrometry places the source near a galaxy detected in the near-IR and mid-IR bands. This is the only galaxy detected within 15$''$ of the DSFG and is a lens candidate.

\item \mbox{ACT-S J0038$-$0022} is located near a complex of apparently unassociated optical/IR sources. The brightest optical source is classified by SDSS as a star. Several galaxies are nearby, including one only detected in the mid-IR. 

\item \mbox{ACT-S J0039+0024} has accurate astrometry placing it nearby a lens candidate detected in the near-IR and mid-IR.

\item \mbox{ACT-S J0044+0118} has no clear optical/IR counterpart. 

\item \mbox{ACT-S J0045$-$0001} is located 20$''$ north of a galaxy cluster bright in the optical/IR. The spectroscopic redshift from SALT follow-up of this cluster is $z=0.234$.

\item \mbox{ACT-S J0107+0001} ($z=3.332$) has SMA astrometry placing it nearby a candidate lens galaxy only detected in the mid-IR.

\item \mbox{ACT-S J0116$-$0004} has SMA astrometry placing it nearby a lens candidate bright in the optical/IR. The photometric redshift of this candidate from SDSS is $z=0.45$.

\item \mbox{ACT-S J0210+0016} ($z=2.553$) is the lensed DSFG published in \citet{Geach15} with a lens redshift from SDSS of $z=0.202$. This source is discussed further in Section \ref{subsec:results_with_redshift} and Appendix \ref{app:0210}.
\end{itemize}

\section {SED modelling}
\label{sec:modeling}

To fit the FIR SEDs we employ a modified blackbody dust emission model with a power-law distribution in dust temperature.  The size distribution and chemical make-up of the dust grains as well as their locations in the radiation field all lead to different temperatures, which can be approximated by a power-law function \citep{ Dale02,Kovacs10}:
\begin{equation}
f(T)=dM_{\rm d}/dT\propto T^{-\gamma}.
\label{pow_law}
\end{equation}  
In this equation, $M_{\rm d}$ is the dust mass, $f(T)$ is normalized as $ \int_{T_{\rm c}}^{\infty }f(T)dT=1$, and $T_{\rm c}$ is the lowest cut-off temperature of the dust.

From the radiative transfer equation, we can simply define the photon escape probability as \citep{Kovacs10}
\begin{equation}
P=1-e^{-\tau} \label{P_escape}.
\end{equation}
In this equation, the optical depth $\tau$ is given by 
\begin{equation}
\tau=\kappa(\nu_r) \Sigma_{\rm d} =\kappa(\nu_r) \frac{M_{\rm d}}{\pi ({d/ 2})^2},
\label{tau}
\end{equation} 
where $\Sigma_{\rm d}$ is the dust surface mass density, $d$ is the dust emission region diameter and $\kappa(\nu_r)=\kappa_0({\nu_r / \nu_0})^{\beta}$  is the mass attenuation coefficient at rest frame frequency $\nu_r$. We normalize $\kappa$ at $c/ \nu_0 =850$~$\mu {\rm m}$ as $\kappa_0=1.5\,{\rm cm^2\,g^{-1}}$ \citep{Weingartner01,Dunne03} and fix $\beta=2$ throughout. This value of $\beta$ is consistent with constraints obtained by \cite{Magnelli12} from \textit{Herschel}-selected DSFGs. We also tested models with $\beta=1.5$, and for the two sources with spectroscopic redshift measurements we fit $\beta$ as a parameter. The effects of relaxing this assumption are investigated in Section \ref{sec:other_models}. The rest frame frequency is related to observed frequency as $\nu_r=(1+z)\nu$. 
The observed flux density from the component of the disk at temperature $T$ can be modelled as:
\begin{equation}
S_\nu(\nu, T)  = \Omega \,  (1 - e^{-\tau(\nu)})\ \, B_{\nu}(\nu, {T/(1+z)}).
\label{eqn:optically_thick}
\end{equation}
In this equation, the solid angle is $\Omega=\pi ({d\over 2})^2/D_{\rm A}^2$, where $D_{\rm A}$ is the angular diameter distance to the source. The spectral radiance is taken to be the Planck function:
\begin{equation}
B_{\nu}(\nu, {T/(1+z)}) ={2h\over c^2}\ {\nu^3\over \exp({h\nu(1+z) /k_{\rm B} T})-1} . 
\end{equation}
This is the single temperature model. 

Applying the power-law temperature distribution (Equation \ref{pow_law}) to the model, we get 
\begin{equation}
S_{\nu,{\rm multi}}(\nu; z, T_{\rm c}, M_{\rm d}, d) = \int_{T_{\rm c}}^{+\infty }f(T)S_{\nu}(\nu,T)dT\\ \nonumber
\end{equation}
\begin{equation}
 = (\gamma-1)T_{\rm c}^{\gamma-1}\int_{T_{\rm c}} ^{+\infty }S_{\nu}(\nu,T)\,T^{-\gamma}\,dT.
\label{multi_model}
\end{equation}
\noindent In this model, $z$, $M_{\rm d}$, $T_{\rm c}$, and $d$ are the parameters to fit. Considering that the shortest wavelength of our photometric data is 250 \mic, our data cover a limited fraction of the Wien side of the Planck function, where the dust temperature distribution parameter $\gamma$ is best constrained. Therefore a typical value for high-$z$ starburst galaxies \citep{Kovacs10,Magnelli12}, $\gamma = 7.0$, is adopted. 

Additionally, we also consider the effect of the Cosmic Microwave Background (CMB) on the dust emission. \citet{DaCunha13} found that the effect of heating by the CMB can become significant for cooler ($T=20$~K) dust at high redshift ($z>5$). Adapting Equations 13, 14, and 15 from \citet{DaCunha13}, the CMB-modified observed flux of the galaxy should be: 
\begin{equation}
S_{\rm obs} = S_{\nu,{\rm multi}} + % \\ \nonumber
\Omega e^{-\tau(\nu)} {2h\over c^2} {\nu^3\over \exp({h\nu /k_{\rm B} T_{\rm CMB})-1}}
\label{add_S_CMB}
\end{equation}
In this equation, $S_{\nu,{\rm multi}}$ is the pure observed flux coming from the dust model (Equation 6) while the latter term captures the contribution from the CMB. As in previous expressions, the frequency $\nu$ is in the observers frame, and so the CMB temperature $T_{\rm CMB}=2.73$~K is taken at $z=0$. For the source in our sample with the highest inferred redshift (ACT-S J0044+0118) and thus the most affected by the CMB, the change in dust model parameters are a few per cent or less relative to the errors on the parameters (representing sub-per cent shifts in parameter values). Therefore, to simplify the analysis, we use only the power-law temperature dust model (Equation \ref{multi_model}) in our results (Section \ref{sec:results}), and do not include the contribution of the CMB.

If the optical depth is small ($\tau \ll 1$), then the escape probability (Equation \ref{P_escape}) becomes $P \approx  \tau$. In this limit, the single-temperature, modified-blackbody model becomes
\begin{eqnarray}
S_\nu(\nu,T)   & = & \Omega \,  \tau(\nu)\ \, B_{\nu}(\nu, {T/(1+z)}) \\
& = & \kappa(\nu(1+z)) \frac{M_{\rm d}}{D_{\rm A}^2}B_{\nu}(\nu, {T/(1+z)})  \nonumber\\
& = & \kappa_{0} (\nu/\nu_0)^\beta (1+z)^\beta \frac{M_{\rm d}}{D_{\rm A}^2}B_{\nu}(\nu, {T/(1+z)})  \nonumber
\label{optically_thin}
\end{eqnarray}
It is noteworthy that this optically thin assumption is frequently employed in SED fitting. However, this assumption is not always suitable, especially for the most luminous and highly-obscured DSFGs in the early Universe. There is growing evidence that high-$z$ DSFGs are characterized by optically thick dust in the submillimetre observing bands \citep{Riechers13, Huang14}. In this study, we model SEDs with Equation 6, using the full functional form for the escape probability.

\subsection{Likelihood analysis}

We use a Gaussian likelihood function $\mathcal{L}$ to describe the distribution of the observed SED about the true emission model. The log-likelihood is therefore given by
\begin{equation}
-2\ln \mathcal{L} \propto \sum_\nu \frac{(D(\nu) - S_{\rm multi}(\nu; z, T_{\rm c}, M_{\rm d}, d))^2}{\sigma^2(\nu)},
\label{eqn:chisq}
\end{equation}
where the sum is over the different frequency bands of the SED $D(\nu)$ with error $\sigma(\nu)$,
and $S_{\rm multi}$ is given by Equation 6. Generalizing this model to account for the possibility of magnification $\mu$, we discuss results in terms of {\it apparent} dust mass $\log_{10}(\appMd/{\rm M}_\odot)$ and effective diameter $\appd$. According to Bayes' Theorem, the posterior probability of the model parameters ($z$, $T_{\rm c}$, $\log_{10}(\appMd/{\rm M}_\odot)$, $\appd$) is  given by the product of this likelihood function and the prior probabilities on parameters  (Section \ref{priors}). We employ an affine-invariant Markov Chain Monte Carlo (MCMC) algorithm \citep{Foreman-Mackey13} to estimate the marginal posterior distributions of multiple parameters. Specifically, for each run of the MCMC we set up 16 chains and iterate each for 6000 steps, allowing the first 500 steps for burn-in. For each fit, the minimum $\chi^2$ (Equation \ref{eqn:chisq}) is found by performing a conjugate gradient search, starting at the minimum of the MCMC sampling.

\subsection{Derived parameters}

In addition to generating the posterior distributions for the model parameters in Equation \ref{eqn:chisq}, we also generate the posterior distribution for the FIR luminosity $L_{\rm IR}$, which we define as
\begin{eqnarray}
L_{\rm IR} = 4\pi D_L^2\int_{\nu_1}^{\nu_2}S_{\rm obs}(\nu; z, T_{\rm c}, M_{\rm d}, d)d\nu,
\label{L_IR}
\end{eqnarray}
where $D_{\rm L}$ is the luminosity distance. The integral is conventionally taken over the rest frame wavelength range $8-1000$~$\mu$m. We also compute the SFR distribution assuming the galaxies are characterized by a Salpeter initial mass function \citep{Salpeter55} in the range of $0.1-100$~\msun\ at a starburst age of 1~Gyr (since the age of the Universe at $z=4$ is about 1.6 Gyr).
Under these assumptions, the SFR is related to the FIR luminosity  by ${\rm SFR}({\rm M}_{\odot}\,{\rm yr}^{-1})=1.06\times10^{-10}L_{\rm IR}({\rm L}_{\odot})$ \citep{Dwek11}. As with dust mass and effective diameter, we present results for FIR luminosity and SFR in terms of apparent quantities: $\logten(\mu \lir/{\rm L}_\odot)$ and $\mu$SFR.
The third and final derived parameter considered in our analysis is $\tau_{100}$, the optical depth at rest frame wavelength $\lambda=100$\,$\mu$m (Equation 3). We choose $\tau_{100}$ because the spectra of DSFGs  typically peak near 100~$\mu$m in the rest frame.

\subsection{Parameter degeneracies and prior constraints}
\label{priors}

\begin{figure*}
  \centering
 \includegraphics[width=7.0in]{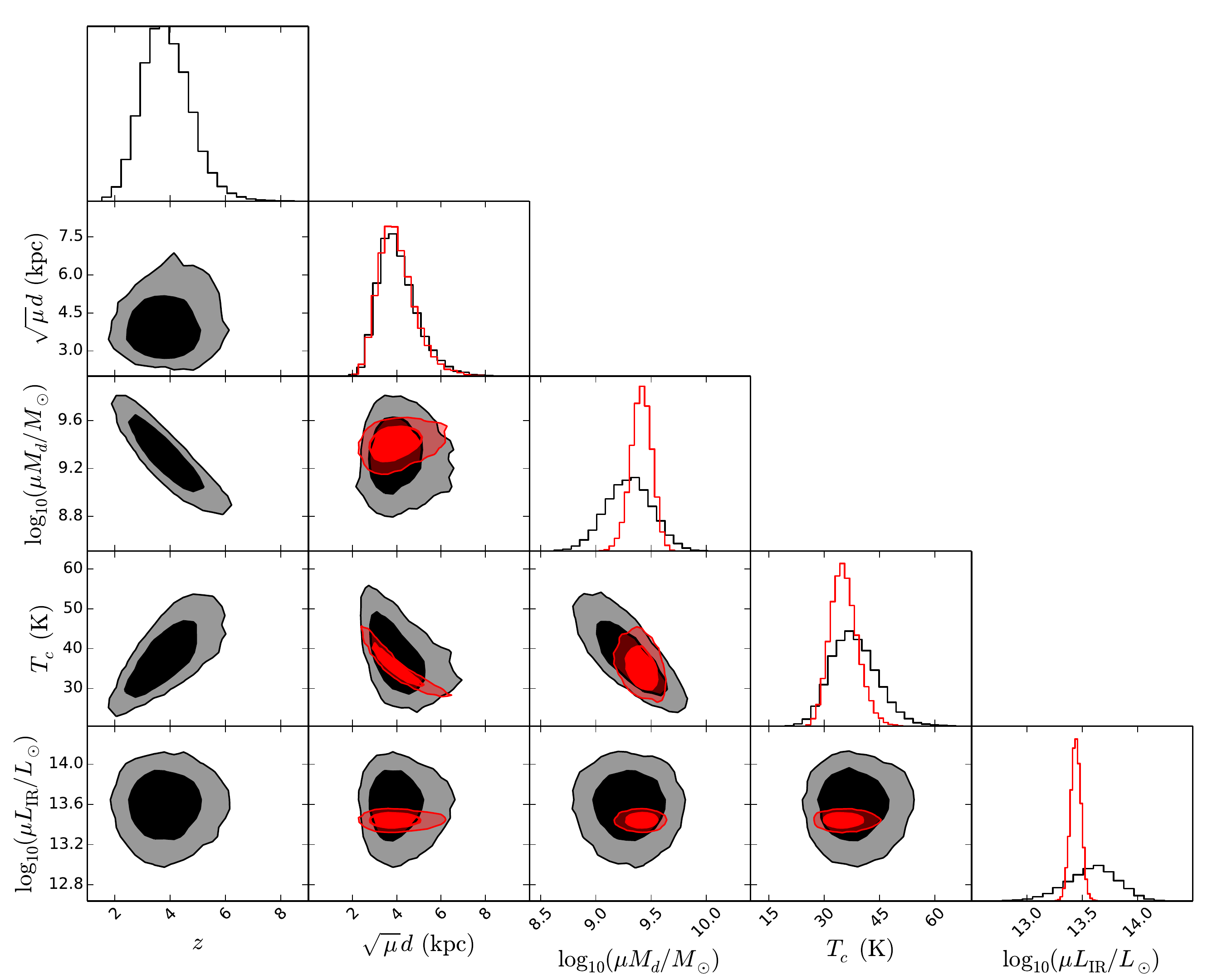} 
 \caption{Posterior distributions of redshift $z$, apparent effective diameter $\appd$, apparent dust mass $\log_{10}(\appMd/{\rm M}_\odot)$, dust temperature distribution cut-off $T_{\rm c}$, and apparent total infrared luminosity $\log_{10}(\applir / {\rm L}_\odot)$ (Equation \ref{L_IR}) for ACT-S J0107+0001. Constraints marginalized over redshift are shown in black, while constraints fixing the redshift to the spectroscopically measured value ($z=3.332$) are shown in red. For the  two-dimensional posterior distributions, the inner and outer contours  bound 68\% and 95\% confidence regions, respectively. Note that the apparent luminosity shows no covariance with other parameters. }
 \label{contours_0107}
\end{figure*}

Degeneracies between model parameters limit the information that can be derived from an SED. The most familiar of these degeneracies is that between the redshift $z$ and temperature $T_{\rm c}$ \citep{Blain99}, which is clear from  the functional form of $B_\nu$. An increase in redshift can be compensated by an increase in temperature. Note, however, that since redshift also enters into the formula for the angular diameter distance $D_{\rm A}$ and $\kappa(\nu)$,  the degeneracy is non-linear.  A second degeneracy exists between the dust mass $M_{\rm d}$ and redshift, which is apparent in the optically thin version of the emission model (Equation 7). These degeneracies can be seen in the two-dimensional posterior probability distributions in Figure \ref{contours_0107}.
Note that the apparent luminosity  shows no covariance with other parameters.

Because of these degeneracies, one cannot obtain interesting constraints on most individual parameters of the modified blackbody dust model using only SED data:  prior information is needed. To make progress we investigate a sample from \citet[W13 hereafter]{weiss13}. The 23 DSFGs in the W13 sample are selected from SPT millimetre-wave data using the criteria that sources must have 220~GHz flux densities in excess of 20~mJy and a spectrum characteristic of dust, which is similar to our selection criteria for the ACT sample. Crucially, these 23 sources have measured spectroscopic redshifts. Following the procedure described in this section, we fit the SEDs of the 23 spectroscopically detected sources in W13. All model parameters  are allowed to vary except for the redshift, which is set to the measured value. 
We use the posterior distributions of $T_{\rm c}$ and $\appMd$ from our analysis of the W13 SEDs as prior probability distributions in fits to our SEDs. The prior distributions are log-normal functions with $T_{\rm c} = 40.7\pm12.2$\,K and $\log_{10}(\appMd/{\rm M}_\odot) = 9.40\pm0.24$. These W13-based priors and sample distributions are plotted with the ACT-sample posterior distributions in Figure \ref{ACT-hist}.\footnote{\citet{Strandet16} have recently extended the W13 sample from SPT. Given the consistency between the original and extended samples, we would not expect the conclusions of this paper to change given the new data.} In terms of these priors, one exception is the case of ACT-S J0210+0016; this source is so bright compared to the other ACT and W13 sources that the goodness-of-fit suffered with the prior on $\appMd$. We therefore use a uniform prior for $\appMd$ when modelling ACT-S J0210+0016.
We note that the range of parameters imposed by these priors (e.g., 28~K~$<T<53$~K), while motivated by the similarity of the ACT and SPT selection, are broad and not exclusive of other results, such as those from \textit{Herschel} and \textit{Planck} \citep{Magnelli12, Bussmann13,Canameras15,Harrington16}.

\begin{table*}
  \centering
  \footnotesize
  \begin{tabular}{@{}cccccccccc@{}}
  \hline
  ID & $z$ &$T_{\rm c}$ & $\log_{10}(\, \mu M_{\rm d} / {\rm M}_{\odot})$  & $\sqrt{\mu} \,d$ &$ \log_{10}(\mu L_{\rm IR} / {\rm L}_{\odot})$&$\mu {\rm SFR}$ &$\tau_{100}$&$\chi^2 / N_{\rm dof}$\\ 
	  &	& [K]		&       &	[kpc]& &[${\rm M}_{\odot}{\rm yr}^{-1}$]& & &\\\hline
ACT-S J0011$-$0018	&$3.3^{+0.8}_{-0.7}$		&$45.9^{+8.5}_{-7.5}$	&$ 9.45\pm0.20$	        &$2.9\pm0.4$    	   	&$13.67^{+0.22}_{-0.24}$ &$5000^{+3200}_{-2100}$       &$10.0^{+6.1}_{-3.7}$ &2.22/2 \\
ACT-S J0022$-$0155	&$4.4^{+1.0}_{-0.8}$	    &$42.6^{+7.1}_{-5.8}$	&$9.36\pm0.20$          &$4.1^{+0.6}_{-0.5}$ 	&$13.83^{+0.20}_{-0.21}$ &$7200^{+4200}_{-2700}$       &$4.0^{+2.3}_{-1.5}$   &0.77/1	\\
ACT-S J0038$-$0022	&$4.3^{+0.9}_{-0.8}$		&$45.5^{+6.9}_{-6.4}$	&$9.35\pm0.20$ 		    &$3.3\pm0.5$	        &$13.87^{+0.18}_{-0.21}$ &$7900^{+4200}_{-3000}$ 	&$4.6^{+2.8}_{-1.7}$ &0.79/2  \\
ACT-S J0039+0024	&$3.3^{+0.8}_{-0.7}$		&$46.8^{+7.9}_{-7.2}$	&$9.35\pm0.20$          &$3.3^{+0.4}_{-0.3}$	&$13.83^{+0.21}_{-0.25}$ &$7300^{+4500}_{-3100}$ 	&$5.8^{+3.4}_{-2.0}$ &1.62/2  \\
ACT-S J0044+0118	&$4.9^{+1.0}_{-0.8}$		&$47.2^{+7.2}_{-6.3}$	&$9.42\pm0.20$          &$4.9\pm0.6$            &$14.17^{+0.18}_{-0.19}$ &$15500^{+7800}_{-5400}$	&$3.2^{+1.7}_{-1.1}$ &1.07/2 \\
ACT-S J0045$-$0001	&$4.6^{+1.0}_{-0.8}$		&$39.9^{+6.4}_{-5.8}$	&$9.39\pm0.21$		&$4.4^{+1.0}_{-0.8}$ 		&$13.67^{+0.202}_{-0.24}$ &$5000^{+3200}_{-2100}$      &$3.6^{+2.7}_{-1.7}$ &0.82/2\\
ACT-S J0107+0001	&3.332				&$35.1^{+3.6}_{-3.5}$	&$9.42^{+0.08}_{-0.09}$ &$3.9^{+0.9}_{-0.7}$ 		    &$13.45\pm0.04$ 		&$3000\pm300$	 	&$4.8^{+2.3}_{-1.6}$&3.83/3\\
ACT-S J0107+0001& $3.8^{+0.9}_{-0.8}$          &$37.2^{+6.4}_{-5.4}$   &$9.31\pm0.20$  &$4.0^{+1.0}_{-0.8}$ 		    &$13.59^{+0.21}_{-0.24}$ &$4200^{+2600}_{-1800}$	        &$3.8^{+3.0}_{-1.8}$ &3.82/2\\
ACT-S J0116$-$0004	&$3.9^{+0.9}_{-0.8}$		&$40.7^{+6.7}_{-5.8}$	&$9.37\pm0.21$		&$5.3^{+1.5}_{-0.9}$ 		&$13.98^{+0.21}_{-0.23}$ &$10000^{+6100}_{-4100}$  &$2.4^{+1.7}_{-1.2}$	&0.50/2\\
ACT-S J0210+0016  	&2.553				        &$42.3\pm2.2$    	&$10.02\pm0.03$		&$7.4^{+0.7}_{-0.6}$ 	    	&$14.34\pm0.03$        &$23200^{+1400}_{-1300}$		&$5.5\pm0.8$&0.81/3\\
ACT-S J0210+0016   &$2.7^{+1.1}_{-0.9}$	   &$43.8^{+12.4}_{-9.8}$  &$9.97^{+0.28}_{-0.31}$  &$7.2^{+0.8}_{-0.7}$        &$14.40^{+0.36}_{-0.40}$  &$26700^{+34200}_{-16200}$  &$5.0^{+3.9}_{+2.2}$ &0.80/2 \\
\hline
ACT Sample    &$4.1^{+1.1}_{-1.0}$              &$43.2^{+8.2}_{-7.2}$   &$9.40^{+0.27}_{-0.22}$ &$4.2^{+1.7}_{-1.0}$ & $13.86^{+0.33}_{-0.30}$  &$7600^{+8600}_{-3900} $&$4.2^{+3.7}_{-1.9}$&12.4/17\\
\hline
\end{tabular}
 \caption{Results from modelling DSFG SED data with a modified blackbody spectrum with a power law dust temperature distribution and without the assumption of optically thin emission (Equation \ref{multi_model}). Properties are  given for individual galaxies detected by ACT at 218~GHz and for the ensemble ACT sample. Results for ACT-S J0107+0001 and ACT-S J0210+0016 are given with and without fixed spectroscopic redshifts. We give the median and the 16th and 84th percentiles for the posterior distribution for each parameter.}
 \label{ACTtable_2}
\end{table*}

%================================================================

\section{Results and discussion}
\label{sec:results}

\subsection{Sample properties}
\label{sec:sampleproperties}

\begin{figure*}
  \centering
  \includegraphics[width=.9\linewidth]{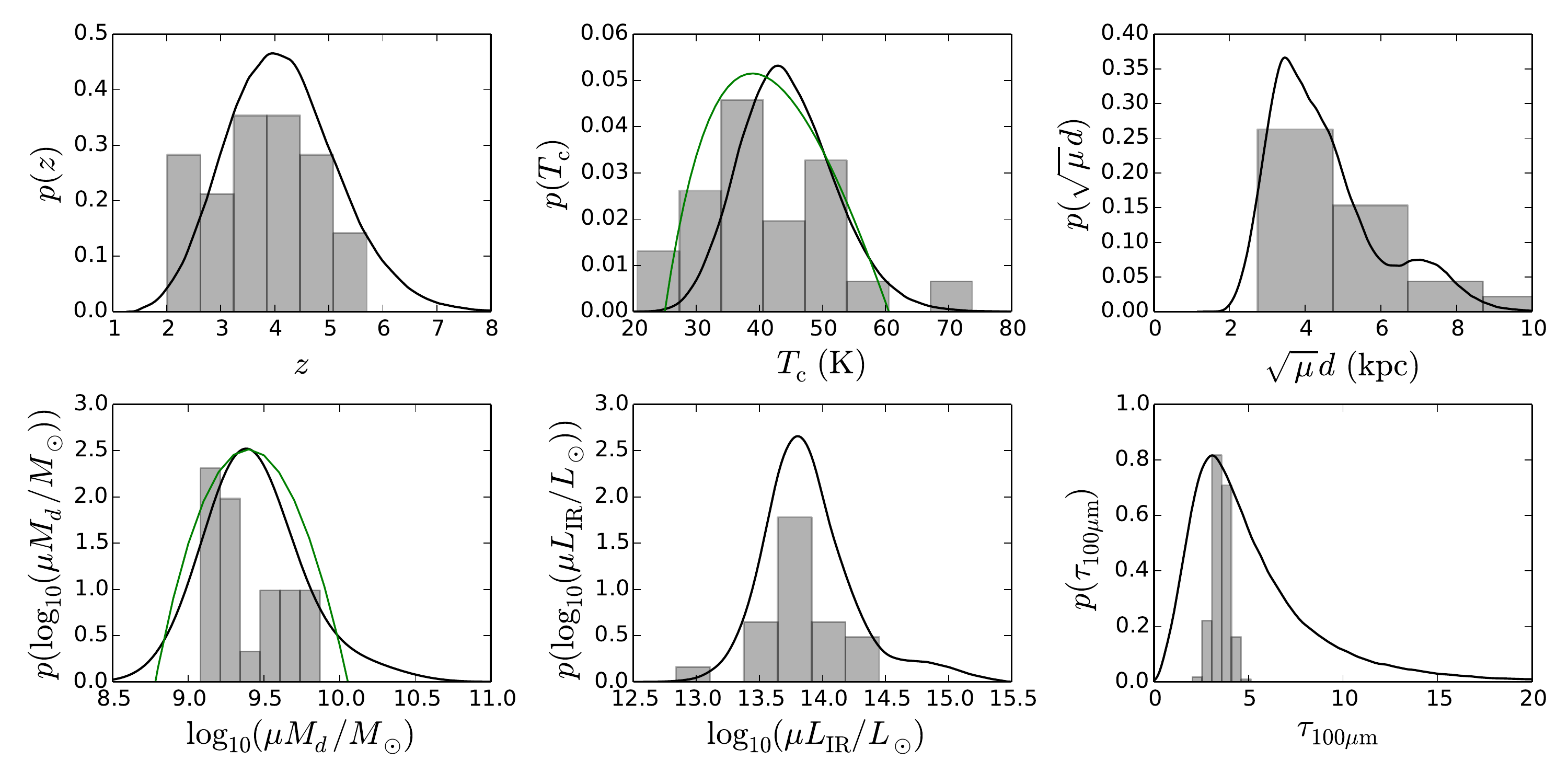}
  \caption{Posterior distributions of model parameters for the ACT sample (black curves). Each distribution is the average of the parameter distributions obtained for individual sources.  Percentiles for these sample distributions are given in Table \ref{ACTtable_2}. The integral of each probability distribution is unit normalized.  High value bump/tails of apparent effective diameter, dust mass and luminosity are due to the highly magnified source ACT-S J0210+0016. For comparison, we also plot the parameter distributions (histograms) for the W13 sample. The W13 redshifts are spectroscopically measured, and the other parameters derive from our fits of the fiducial model (Equation \ref{multi_model}) to SED data in W13, fixing the redshift to the spectroscopic value. The derived prior distributions for $T_{\rm c}$ and $\logten{(\appMd/{\rm M}_\odot)}$ used to fit our data are shown with green curves.}
  \label{ACT-hist}
\end{figure*}

Following the modelling procedure described above, we fit the modified blackbody model with power-law temperature distribution (Equation 6) to the ACT and \textit{Herschel} data (Table \ref{ACTtable}). 
We show the results of the 4-parameter ($z$, $T_{\rm c}$, $\log_{10}(\mu {M}_{\rm d}/{\rm M_\odot})$, $\sqrt{\mu}d$)  fitting  of our sample in Table \ref{ACTtable_2} and Figure \ref{ACT-hist}. We present the results for the nine individual sources and the whole sample as well. The distributions of physical properties for the sample are the averages of the posterior distributions of the individual galaxies.

The total $\chi^2$  of the fits is 12.4 for 17 degrees of freedom, giving a {\it p}-value for the sample of 0.775. Therefore the fiducial model provides an acceptable fit to the data. As modelled, the ACT-selected sample has a  redshift of $z=4.1^{+1.1}_{-1.0}$ and an apparent diameter of $\sqrt{\mu}d=4.2^{+1.7}_{-1.0}$~kpc. These values are the medians of the posterior distributions in Figure \ref{ACT-hist} with error bounds given by the 16th and 84th percentiles. The cutoff temperature of the sample is $T_c=43.2^{+8.2}_{-7.2}$~K and the dust mass is $\log_{10}(\, \mu M_{\rm d} / {\rm M}_{\odot})=9.40^{+0.27}_{-0.22}$. The temperature and dust mass were constrained by priors (Section \ref{priors}). The temperature posterior distribution's median is a bit higher and its width a bit narrower in comparison to the prior. The dust mass posterior distribution is comparable to the prior, with some extra width due to the effect of the highly magnified source ACT-S J0210+0016. See Figure \ref{ACT-hist} where the priors are plotted with the posteriors. The derived median apparent FIR luminosity is $\logten(\mu \lir/{\rm L}_\odot)=13.86^{+0.33}_{-0.30}$ with corresponding  $\mu$SFR~$=7600^{+8600}_{-3900}$~${\rm M}_\odot{\rm yr}^{-1}$. The derived median optical depth is $\tau_{100}=4.2^{+3.7}_{-1.9}$ at 100\,$\mu$m. High value tails of the distributions for the apparent effective diameter, dust mass and luminosity are due to the highly magnified source ACT-S J0210+0016.

Derived from an integral of the apparent specific luminosity over the entire infrared range, $\log_{10}(\mu L_{\rm IR} / {\rm L}_{\odot})$ is one of the best quantities determined from fluxes. While other model parameters have significant covariance, the apparent total luminosity is relatively independent of the other parameters. To illustrate this point, we show, in the last row of Figure \ref{contours_0107}, the two-dimensional posterior distributions for $\log_{10}(\mu L_{\rm IR} / {\rm L}_{\odot})$ and other parameters. The symmetry of these distributions indicates that $\log_{10}(\mu L_{\rm IR} / {\rm L}_{\odot})$ is not strongly correlated with the other parameters. Some intuition for why this is can be developed by considering the other parameter degeneracies: if $z$ and $T_c$ decrease, the overall amplitudes of the fluxes predicted by the model decrease along with the the apparent total luminosity. To match the flux density data, the dust mass and size of the galaxy increase, compensating the loss in $\log_{10}(\mu L_{\rm IR} / {\rm L}_{\odot})$ for the decrease in $z$ and $T_c$. As a result, $\log_{10}(\mu L_{\rm IR} / {\rm L}_{\odot})$ is relatively stable in the face of other parameter shifts. The total apparent luminosity is also one of the more model-independent quantities (Section \ref{sec:other_models} and Appendix \ref{SEDfitting}).

\subsection{Constraints with redshift information}
\label{subsec:results_with_redshift}

As discussed in Section \ref{sec:data_other} and Appendix \ref{app:0210}, we have spectroscopic redshift measurements for ACT-S J0107+0001 and ACT-S J0210+0016. Table \ref{ACTtable_2} shows the results of fitting these sources with the redshifts  fixed to the spectroscopic values. The resulting parameter constraints are seen to be consistent with and (unsurprisingly) tighter than constraints from fitting the data with a flat, unbounded prior on the redshifts. (See also Figure \ref{contours_0107}.)
\citet{Geach15}  have also studied ACT-S J0210+0016. Using a single-temperature, modified-blackbody SED model and the known redshift, they find that $T = 38 \pm 4$ K and $\beta = 1.8\pm 0.4$, results that are in agreement with our estimate of $T_{\rm c} = 42.3\pm2.2$ K and the assumption $\beta = 2.0$. However, \citet{Geach15} estimate the apparent FIR luminosity as $\applir = (1.3 \pm 0.1) \times 10^{14}$\,${\rm L}_{\odot}$, while our fitting gives $\applir = 2.2^{+0.2}_{-0.1}\times 10^{14}$\,${\rm L}_{\odot}$. The difference arises because the single-temperature model fails to catch the mid-IR excess of the SED at the Wien side, which leads to the underestimation of the total infrared luminosity \citep[][]{Kovacs10,Magnelli12}. %See also Appendix \ref{app:spt}.
Through lens modelling, \citet{Geach15} estimate the lens magnification $\mu \sim$ $11$--$13$.
Applying this magnification to our results, the intrinsic properties of ACT-S J0210+0016 are $M_{\rm d} = 1.20^{+0.15}_{-0.11}\times 10^{9}$~${\rm M}_{\odot}$, $d =  2.1^{+0.3}_{-0.2}$\,kpc and $L_{\rm IR} = 1.8^{+0.3}_{-0.2} \times 10^{13}$\,${\rm L}_{\odot}$, which reveals that it is also intrinsically very luminous. Additionally, the emission region diameter is consistent with  typical intrinsic diameters of SMGs, which are 1 -- 3 kpc \citep[e.g.][]{Kovacs10,Magnelli12,Hezaveh13, Riechers13,Simpson15,Spilker16}.

\subsection{Other models} 
\label{sec:other_models}
 
For comparison, an additional three models, including single-temperature models with and without the assumption of optically thin dust and a power-law temperature distribution model with optically thin dust, are fit to the SED data in Appendix \ref{SEDfitting}. These are included to facilitate straightforward comparison to other works that use these models and to highlight and quantify the systematic errors intrinsic to SED modeling. 

The single-temperature, optically thin model fits the data with total $\chi^2$ of 136.4 for 26 degrees of freedom (with a vanishing $p$-value of $10^{-35}$). Note that one more degree of freedom per source is included in this model because the optically thin assumption eliminates one parameter, which we have chosen to be $d$ (Equation \ref{optically_thin}). Thus the fiducial model (Equation \ref{multi_model}) is strongly preferred over the simpler single-temperature, optically thin model.

The model with a power-law temperature distribution and optically thin dust gives a $\chi^2$ of 34.4 for 26 degrees of freedom (p-value of 0.125). Therefore, while formally worse than our fiducial model, this fit is still acceptable. One challenge to taking the results at face value, however, is the exceptionally high apparent luminosities corresponding to apparent star formation rates in the tens of thousands of solar masses per year (with a whopping $2.5\times10^5$~M$_\odot$yr$^{-1}$ for ACT-S J0210+0016). The corresponding high redshifts given by this model (with a sample median redshift of $z=6.8$) would imply a truly exceptional source population -- one that has not been observed. Additionally this model's redshifts for ACT-S J0107+0001 and ACT-S J0210+0016 are twice those measured by  CO spectroscopy. The extraordinary characteristics of this model suggest that again the fiducial model is preferred. Notably, these  investigations disfavor both models with optically thin dust.

The single-temperature model without the assumption of optically thin dust gives a $\chi^2$ of 17.3 for 17 degrees of freedom (p-value of 0.434). Therefore, in terms of goodness of fit, this model is on par with our fiducial model. The data cannot distinguish between models based on whether they assume a single dust temperature  or power-law dust temperature distribution. The comparison of the parameter constraints of this model with those of the fiducial model is non-trivial. We find the median sample redshift reduces to $z=3.3$ from $z=4.1$. The median dust temperature for the single-temperature model is $T=52$~K while the lower cutoff temperature of the fiducial model is $T_c=43$~K. It is expected that the $T_c$, as the minimum temperature of a distribution, would be lower than the single-temperature $T$. Finally, the median sample opacity increases to $\tau_{100}=5.9$ from $\tau_{100}=4.2$ for the fiducial model. The increase in opacity reduces the frequency of the peak in the single temperature model, compensating for the lower redshift. The median sample luminosity of the single-temperature model is lower by 0.26 dex relative to the fiducial model. Referring to Figure \ref{SEDs_appendix}, this is a generic consequence of the fact that hotter dust in the power-law temperature distribution extends the composite modified black-body spectrum to higher frequencies. We note that all of the parameter shifts are within the one-sigma errors of the two model fits: for these data, the systematic uncertainties from model choice are comparable to statistical uncertainties in a given model. In the end, the conclusion that the average source in the ACT-selected sample is many times more luminous than a typical ULIRG, located at a redshift beyond the peak in the cosmic star formation rate history, and characterized by relatively hot ($T\sim50$~K) and not optically thin dust is independent of whether the model assumes single-temperature dust or dust with a power-law distribution of temperatures. We have chosen the fiducial model to have  a power-law temperature distribution instead of a single temperature based on physical arguments and studies at shorter wavelengths that favor the power-law distribution \citep[e.g.][]{Kovacs10}.  As can be seen in  Figure \ref{SEDs_appendix}, data with wavelength shortward 250\,$\mu$m, on the Wien side of the modified black-body spectrum, can distinguish between models with a single-temperature and a power-law temperature distribution. 

For the fiducial model we also considered setting the emissivity parameter to $\beta=1.5$ (instead of the fiducial $\beta=2.0$). Compared to the fiducial model, the $\beta=1.5$ model produced a formally worse (but still acceptable) fit with $\chi^2=25.0$ for 17 degrees of freedom giving a $p$-value of 0.096. For this value of $\beta$, the parameters $z$, $T_{\rm c}$, and $\appd$ increased by one standard deviation whereas $\logten{\appMd}$ did not change significantly. Finally, for the sources with spectroscopic redshift measurements, treating $\beta$ as a parameter (while fixing $z$ to its measured value) gave $\beta=2.3^{+0.5}_{-0.6}$ and $\beta=2.2^{+0.4}_{-0.3}$ for ACT-S J0210+0016 and ACT-S J0107+0001, respectively, and results similar to the fiducial $\beta=2.0$ model for the other fit parameters.

\subsection{Discussion}

\begin{figure*}
 \includegraphics[width=6.5in]{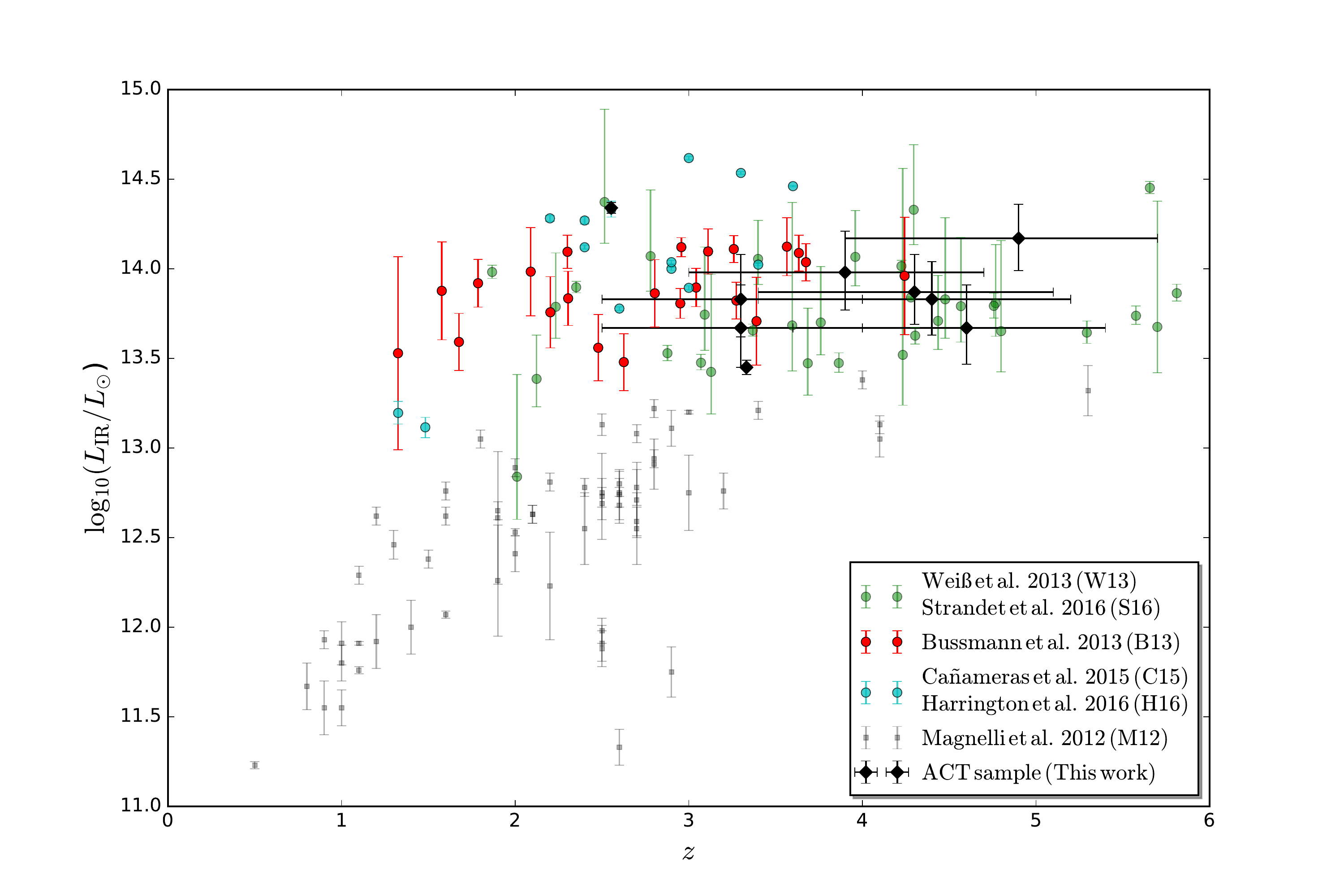} 
 \caption{Apparent $L_{\rm IR}$ versus $z$ for DSFGs from millimetre and submillimetre-selected samples. The sources of B13, W13/S16, and C15/H16 samples are primarily lensed, while the M12 sources are unlensed. Some data (especially for C15/H16) have errorbars that are smaller than the symbols due to the precision of the corresponding photometry. At the same redshift, the ACT sample has apparent total infrared luminosities comparable to the lensed samples and $5$--$10$ times  those of the 
 unlensed M12 samples, indicating the presence of strong lensing or multiple unresolved sources. The ACT sample has a higher median redshift than the M12 and B13 samples, which are selected at higher frequencies. }
 \label{LIR_z}
\end{figure*}

Figure \ref{LIR_z} compares the ACT sample to others found in the literature in terms of apparent total infrared luminosity and redshift. We compare to the samples of W13 and \citet{Strandet16} (SPT 220~GHz (1.4\,mm), hereafter W13/S16), the samples of \citet{Canameras15} and \cite{Harrington16} ({\it Planck}-\textit{Herschel} SPIRE 250-850~$\mu$m; hereafter C15/H16), 
the sample of \citet{Bussmann13} (\textit{Herschel} SPIRE 250\,$\mu$m, 350\,$\mu$m,  500\,$\mu$m, with emphasis on 500\,$\mu$m; hereafter B13), and the sample of \citet{Magnelli12} (\textit{Herschel} PACS/SPIRE $100$--$500$\,$\mu$m; hereafter M12).
There is evidence that most sources in the B13, W13/S16, and C15/H16 samples are lensed, whereas the sources in the M12 sample are primarily unlensed. The apparent total infrared luminosities plotted in Figure \ref{LIR_z} for M12, W13/S16, and C15/H16 were obtained by fitting the fiducial multi-temperature SED model (Section \ref{sec:modeling}, Equation \ref{multi_model}) to data provided in those papers, while those for B13 were obtained with the single-temperature optically thin model. (B13 did not provide SED data.) The B13 model will yield slightly different apparent total infrared luminosities. This difference is small enough to be ignored in this comparison. For example, although the single-temperature model with optically thin dust is a poor fit to our data, this model's result for apparent luminosity $\log_{10}(\applir /{\rm L}_\odot)=13.87^{+0.40}_{-0.29}$ (Table~\ref{1Tthin_results})  is close to that of our fiducial model $\log_{10}(\applir /{\rm L}_\odot)=13.86^{+0.33}_{-0.30}$ (Table~\ref{ACTtable_2}).
The apparent total infrared luminosities of the ACT sample are comparable to those of the lensed samples of B13, W13/S16, and C15/H16, which are characterized by magnifications of $\mu\sim10$. As discussed in Section \ref{sec:sampleproperties} the apparent luminosity is a relatively robust parameter in terms of model choice and parameter degeneracy. Also the temperature distribution of our sample falls within the broad priors chosen and is in agreement with the lensed samples, suggesting that the high apparent luminosity of our sample does not come primarily from extra dust heating. Taken together these arguments offer evidence that the ACT-selected sample is lensed similarly to the B13, W13/S16, and C15/H16 samples.

Comparison of apparent effective diameters of these sources to direct size measurements supports a similar conclusion. \citet{Simpson15} present ALMA observations of 23 SCUBA-2-selected SMGs with a median physical half-light diameter of $2.4\pm 0.2$\,kpc, while \citet{Ikarashi15} show ALMA observations of 13 AzTEC-selected SMGs with a median  physical half-light diameter of $1.34^{+0.26}_{-0.28}$\,kpc.  ALMA observations of four SPT-selected lensed SMGs give a mean physical half-light diameter of 2.14\,kpc \citep{Hezaveh13}. This measurement is consistent with a recent lensing analysis of a significantly expanded SPT-selected DSFG sample \citep{Spilker16}. These high-resolution ALMA observations constrain the far-infrared sizes of the sources to be $1.0$--$2.5$~kpc. Earlier observations of the physical sizes of SMGs by CO detection and 1.4\,GHz imaging suggest larger sizes \citep[e.g.,][]{Tacconi06,Biggs08,Younger08}. However, \citet{Simpson15} point out that the submillimetre sizes are consistent with resolved $^{12}$CO detections, while the sizes derived from 1.4\,GHz imaging are about two times larger because of the cosmic ray diffusion, which can explain the results before higher frequency
observations at ALMA were possible \citep{Chapman04,Tacconi06,Biggs08,Younger08}. Similarly, \citet{Ikarashi15} reveal that the $^{12}$CO detected sizes and the 1.4\,GHz imaging sizes of similar sources are greater than their submillimetre sizes as well. Furthermore, observations of local galaxies also show the submillimetre sizes are smaller than the CO detected sizes \citep[e.g.][]{Sakamoto06,Sakamoto08,Wilson08} and the 1.4\,GHz continuum sizes \citep[e.g.,][]{Elbaz11}. Our photometrically derived $\appd$ is best compared to the submillimetre continuum sizes. With a median apparent effective diameter of  $4.2^{+1.7}_{-1.0}$\,kpc, the  $\appd$ of our sample is $1$--$6$ times the observed intrinsic diameters ($1.0-2.5$~kpc). Lensing (or multiplicity) increases the apparent effective size of a source, so this comparison favors a lensing (or multiplicity) interpretation for the ACT-selected sources.

Additionally, \citet{Wardlow13}  present a  statistical lensing  model based on the flux density at a wavelength of 500\,$\mu$m ($S_{\rm 500 \mu {\rm m} }$, corresponding to a frequency of 600\,GHz) and derive the distribution of lensing magnification $\mu$ as a function of $S_{\rm 500 \mu {\rm m} }$. The median $S_{\rm 500\mu {\rm m}}$ of our ACT sample is 91 mJy. As shown in Figure 8 of \cite{Wardlow13}, the expected magnification at this flux density is $\mu\sim8$, which is consistent with the evidence for strong lensing based on comparisons of luminosity and size.

It is expected that the ACT sample, selected at 218\,GHz (1.4\,mm), will have a higher median redshift than samples selected at higher frequency \citep{weiss13,Symeonidis11,Bethermin15}: due to the negative K-correction associated with the Rayleigh-Jeans tail of the Planck spectrum, the highest-redshift sources remain bright at 218\,GHz. %while their higher frequency fluxes might get dim. 
In contrast, these same sources will dim at higher frequency as the peak in their thermal spectrum shifts \citep[e.g.,][]{Casey14}. This prediction is consistent with the results of this work. Our sample shows a comparable redshift ($z\sim4$) to the 220~GHz-selected sample of W13/S16. At shorter wavelengths B13, selected to be bright at 500~$\mu$m, has $z\sim3$, and M12, selected over all {\it Herschel} PACS/SPIRE bands, has $z\sim 2.5$. Notably, recent studies of {\it Herschel}-SPIRE  ``red'' sources, selected with maximum flux at $500$~$\mu$m, yield substantial, largely unlensed samples with high redshifts $z>3.5$ \citep{Asboth16,Nayyeri16}. In fact, we have used the catalog from \citet{Asboth16} for part of our {\it Herschel} dataset (Section \ref{sec:herschel}). 

The optical depth found in this study ($\tau_{100}=4.2^{+3.6}_{-1.9}$) disfavors the assumption of optically thin dust, a conclusion that is robust against model choice (Section \ref{sec:other_models}). This result is consistent with other observations at high-$z$, where galaxies are heavily enshrouded by dust \citep{Bussmann13,Riechers14}, and the most intense starburst galaxies in the local Universe \citep[e.g.,][]{Wilson14}.

\section{Conclusions}
\label{sec:conclusions}

We have presented nine ACT 218~GHz-selected DSFGs with multi-wavelength detections from 250 $\mu {\rm m}$ to 2 mm. The millimetre/submillimetre photometry has been modelled with a modified blackbody spectrum with power-law dust temperature distribution and without the assumption of optically thin dust. We have assumed broad priors on dust temperature and mass consistent with the results of a range of analogous millimetre/submillimetre studies. Thus modelled, the ACT sample has a redshift distribution with median $z=4.1^{+1.1}_{-1.0}$, which is consistent with a 218\,GHz selection and higher than the redshifts characteristic of samples selected at shorter wavelengths.
The sample has an apparent total infrared luminosity $\log_{10}(\applir/{\rm L}_\odot) = 13.86^{+0.33}_{-0.30}$  and an apparent effective diameter $\appd=4.2^{+1.7}_{-1.0}$ kpc, values  indicative of  strong lensing and/or  multiple unresolved sources.
The sample's characteristic optical depth is $\tau_{100}=4.2^{+3.6}_{-1.9}$ at 100\,$\mu$m. 
We have considered a range of other models and find that models without the assumption of  optically thin dust are preferred. These results are in broad agreement with other studies of millimetre/submillimetre-selected, lensed, high-redshift galaxies \citep{Wardlow13,Bussmann13,weiss13,Strandet16,Canameras15,Harrington16}.

This is the first publication devoted to the study of ACT-selected DSFGs. An ongoing multi-wavelength observing campaign on the parent sample will yield insights into galaxy formation at high redshift through studies of the DSFGs \cite[e.g.,][]{Swinbank10,Bothwell13} and into the structure of dark matter haloes through studies of their lenses \cite[e.g.,][]{fadely/keeton:2012,Hezaveh13a,Hezaveh16}. These studies will set the stage for work on larger ACT-selected samples: a new generation ACT instrument (Advanced ACTPol) is beginning an extragalactic survey of half the sky at three times the depth of the present sample at 1.4~mm wavelength with complementary data at 2~mm and lower frequencies \citep{Henderson15}. Model extrapolations \cite[e.g.,][]{Bethermin12,Cai13} to such a wide and deep survey imply many thousands of lensed and unlensed DSFGs will be uncovered. We look forward to the new discovery space and enhanced statistical constraints of the future sample.

\section*{Acknowledgements}
We thank our LMT collaborators for permission
to make use of the spectroscopic redshift for ACT-S J0107+0001 in advance of publication. We thank Zhen-Yi Cai for providing model source distributions. AJB acknowledges support from the National Science Foundation though grant AST-0955810. ACT was supported by the U.S. National Science Foundation through awards AST-0408698 and AST-0965625 for the ACT project, as well as awards PHY-0855887 and PHY-1214379. ACT funding was also provided by Princeton University, the University of Pennsylvania, and a Canada Foundation for Innovation (CFI) award to UBC. ACT operates in the Parque Astron\'omico Atacama in northern Chile under the auspices of the Comisi\'on Nacional de Investigaci\'on Cient\'ifica y Tecnol\'ogica de Chile (CONICYT). Computations were performed on the GPC supercomputer at the SciNet HPC Consortium. SciNet is funded by the CFI under the auspices of Compute Canada, the Government of Ontario, the Ontario Research Fund -- Research Excellence; and the University of Toronto. Support for CARMA construction was derived from the Moore and Norris Foundations, the Associates of Caltech, the states of California, Illinois, and Maryland, and the NSF.  CARMA development and operations were supported by the NSF under a cooperative agreement, and by the CARMA partner universities. The National Radio Astronomy Observatory is a facility of the National Science Foundation operated under cooperative agreement by Associated Universities, Inc. We have used optical imaging from SDSS. Funding for the SDSS and SDSS-II has been provided by the Alfred P. Sloan Foundation, the Participating Institutions, the National Science Foundation, the U.S. Department of Energy, the National Aeronautics and Space Administration, the Japanese Monbukagakusho, the Max Planck Society, and the Higher Education Funding Council for England. Funding for SDSS-III has been provided by the Alfred P. Sloan Foundation, the Participating Institutions, the National Science Foundation, and the U.S. Department of Energy Office of Science. Part of our NIR imaging is based on observations obtained as part of the VISTA Hemisphere Survey, ESO Progam, 179.A-2010 (PI: McMahon). We also have used data based on observations obtained with the Apache Point Observatory 3.5-meter telescope, which is owned and operated by the Astrophysical Research Consortium. This publication makes use of data products from the Wide-field Infrared Survey Explorer, which is a joint project of the University of California, Los Angeles, and the Jet Propulsion Laboratory/California Institute of Technology, funded by the National Aeronautics and Space Administration. Some of the observations reported in this paper were obtained with the Southern African Large Telescope (SALT). Finally, we acknowledge the MNRAS reviewer and editor for comments that improved the paper.

\setlength{\bibsep}{0pt}
\bibliography{ACTbib}

\appendix
\section{The Lensed DSFG ACT-S J0210+0016}
\label{app:0210}

As noted in section 4.2, ACT-S\,J0210+0016 has also been observed by
\citet{Geach15}. Our independent program to determine its redshift began on
2013 February 6, with observations using the Zpectrometer cross-correlation
spectrometer \citep{Harris07} and the dual-channel Ka-band correlation
receiver on the Robert C. Byrd Green Bank Telescope (GBT).\footnote{Project
ID = 13A-474} We observed the source's 218\,GHz ACT position (J2000
coordinates $\ang{02;09;41.0}$ and $+00^h15^m57.0^s$) for $25 \times 4$ minute scans,
alternating with an equal number of 4 minute scans at the position of a
different DSFG treated as an effective ``sky'' pointing, and making roughly
hourly visits to the nearby quasar J0217+0144 in order to track changes in
pointing, focus, and system gain. Flux calibration was determined from
contemporaneous observations of 3C48, adopting a Ka-band flux density of
0.86\,Jy at 32.0\,GHz, based on the 2012 fitting function of \cite{Perley13}. Conversion from the
Zpectrometer's native lag data to a spectrum used an internal calibration data
set obtained at the beginning of the observing session.  By taking the
difference of the spectra towards our source and ``sky'' positions, we
eliminated systematic baseline structure due to optical imbalance and obtained
a single difference spectrum with a flat baseline from 25.6 to 37.7\,GHz.
The +0.42\,mJy continuum offset in this difference spectrum means that 
ACT-S\,J0210+0016 is brighter than the DSFG at the ``sky'' position with which
it was paired, although we cannot determine a continuum flux density for
either source individually.  The Zpectrometer has a channel width of 8\,MHz,
but its frequency {\it resolution} corresponds to a sinc function with FWHM
20\,MHz;  \citet{Harris10} provides details on this and other aspects of
Zpectrometer data acquisition and reduction.

\begin{figure*}
  \centering
 \includegraphics[trim=0.5cm 9cm 0.5cm 9cm,width=.55\linewidth]{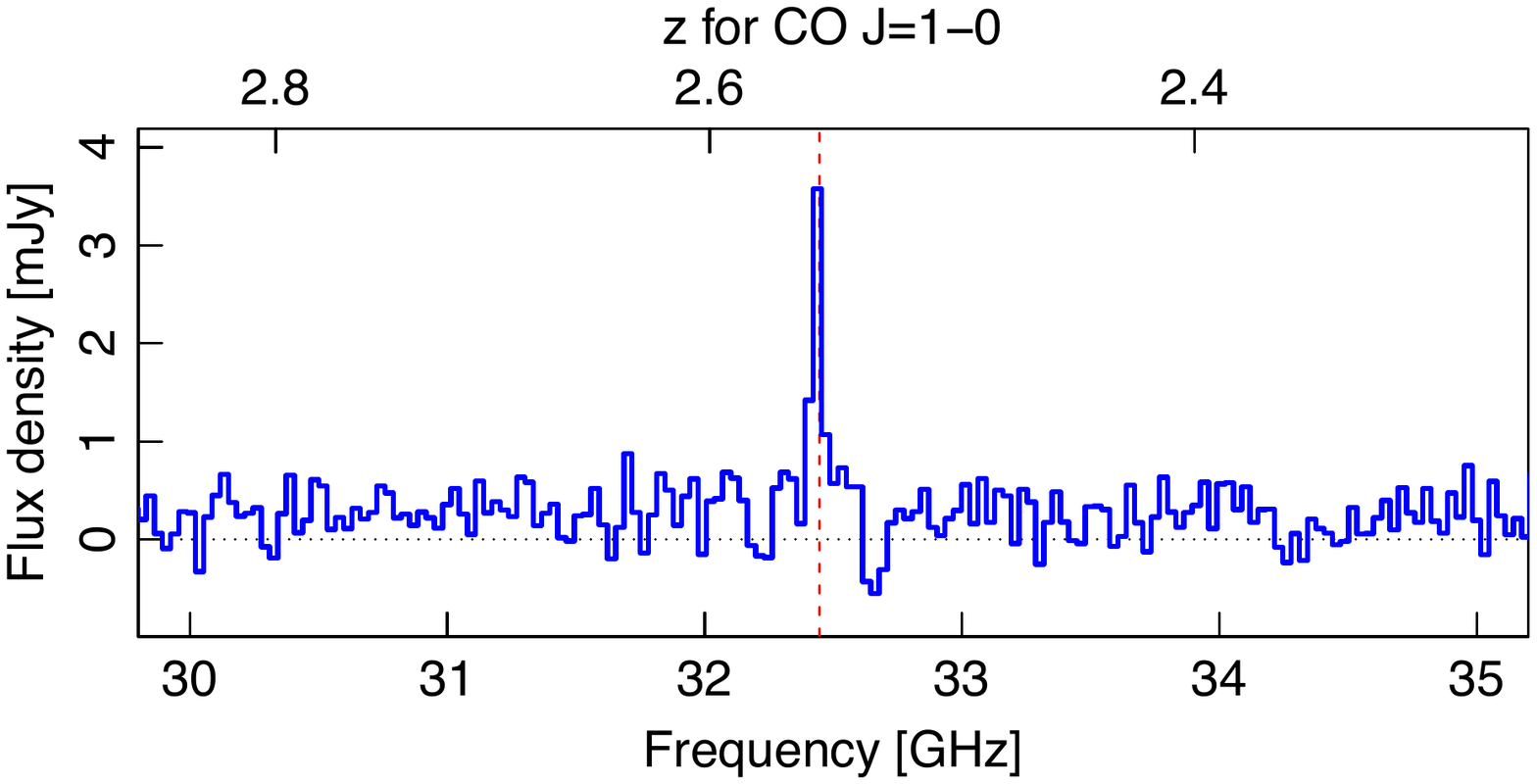} 
 \caption{GBT/Zpectrometer CO(1--0) spectrum of ACT-S\,J0210+0016.}
 \label{X}
\end{figure*}

We clearly detect a positive feature in the difference spectrum (Figure
\ref{X}), at a frequency of $(32.445 \pm 0.001)\,{\rm GHz}$ that corresponds to
a topocentric redshift of $2.55293 \pm 0.00011$ for the CO(1--0) line whose
identification is confirmed below.  The peak flux density is $(5.11 \pm
0.44)\,{\rm mJy}$, and the velocity width of the line is $(259 \pm 28)\,{\rm
  km\,s^{-1}}$, giving a best estimate for the line flux of $(1.41 \pm 0.19)\,{\rm
  Jy\,km\,s^{-1}}$.  For the cosmology adopted in this paper, this line flux
corresponds to an apparent CO(1--0) line luminosity of $\mu L^\prime_{\rm
  CO(1-0)} = (4.48 \pm 0.60) \times 10^{11}\,{\rm K\,km\,s^{-1}\,pc^2}$
 \citep[e.g.][]{Carilli13}. \citet{Harris12} have shown that apparent
$L^\prime_{\rm CO(1-0)}$ and CO(1--0) FWHM velocity width can be used to
estimate a DSFG's lensing magnification to within a factor of around 2 of the
results of a detailed lens model. That paper's scaling relation (Equation 2)
predicts $\mu \approx 33$, higher than the value estimated by \citet{Geach15}
from detailed lens modelling ($\mu \approx$ $11$--$13$) but within the scatter
of the \citet{Harris12} relation.

To confirm the suspected line identification, we also observed
ACT-S\,J0210+0016 for three sessions in 2014 January and February with the
Combined Array for Research in Millimeter-wave Astronomy (CARMA).  CARMA
comprised $6 \times 10.4\,{\rm m}$ telescopes and $9 \times 6.1\,{\rm m}$
telescopes, which during our observation were laid out in its D configuration.
We tuned the 3\,mm receivers to 97.330\,GHz, which would correspond to the
redshifted CO(3--2) line if our assumed identification of the GBT detection were
correct.  We deployed eight spectral windows apiece across the upper and
lower sidebands; each sideband had one narrowband spectral window (248.7\,MHz
with $1.30\,{\rm MHz}$ or $4.0\,{\rm km\,s^{-1}}$ resolution) and
seven wideband spectral windows (487.5\,MHz with
12.5\,MHz resolution). The pointing centre was the ACT 218\,GHz centroid
noted above. Observations of the nearby quasar J0224+069 every 15 minutes
were used for phase and amplitude calibration, 3C84 and 3C454.3 were used for
passband calibration, and Uranus defined the overall flux scale. All data were
reduced using the MIRIAD package \citep{Sault95}.  Following pipeline
pre-flagging and manual phase and amplitude flagging, we were left with a
total observing time of 6.7\,hr.  We imaged the data using the MIRIAD
command {\tt mossdi}, with robust weighting (robust parameter = 2) delivering
a synthesized beam of $5.68\,\arcsec \times 4.29\,\arcsec$ at a
position angle of $\ang{32.4}$. The RMS in each $4.0\,{\rm km\,s^{-1}}$ Hanning-smoothed
channel is $13\,{\rm mJy\,beam^{-1}}$.

\begin{figure*}
\includegraphics[trim=0.5cm 0.5cm 0.5cm 10cm,width=3in]{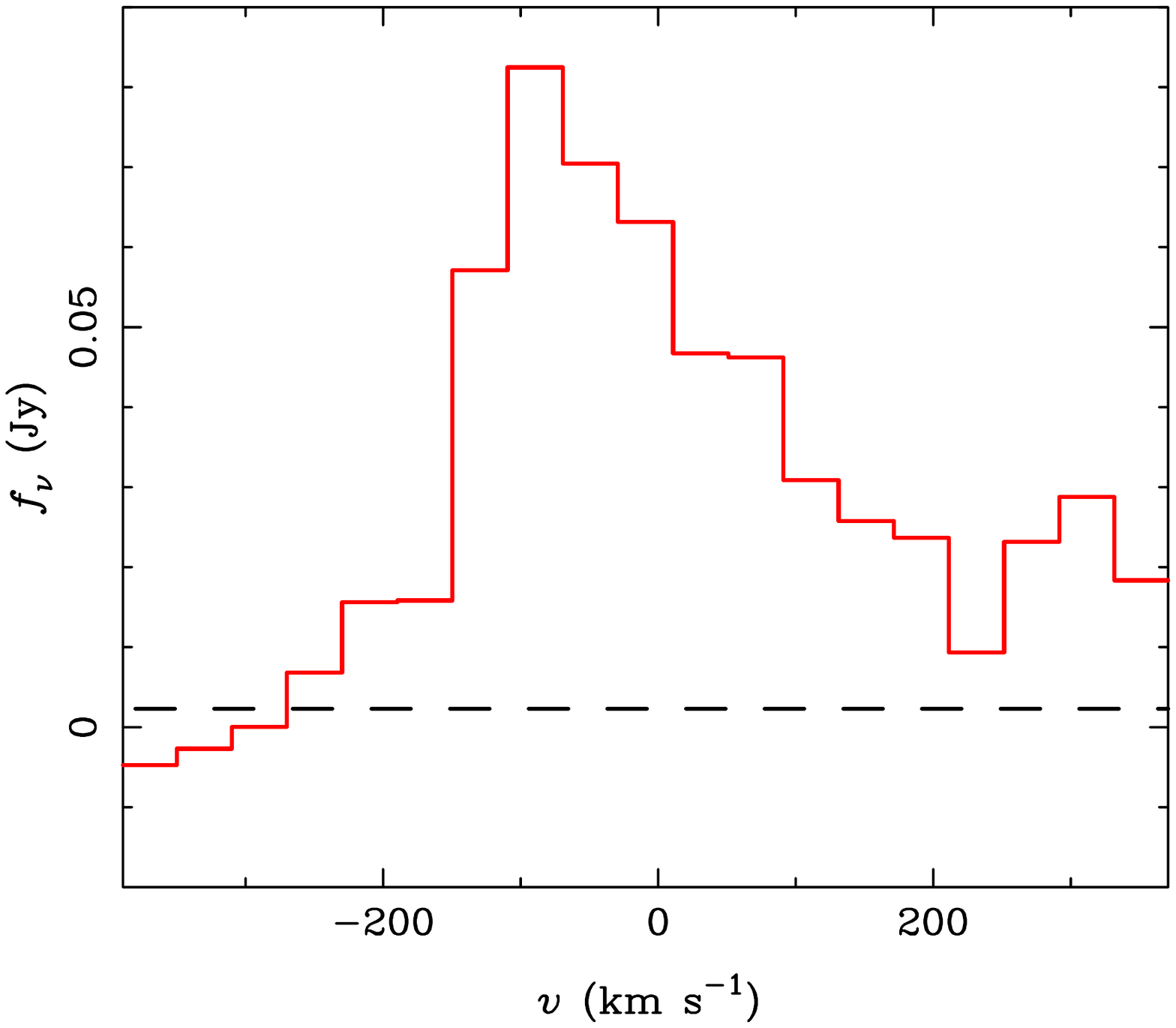} 
 \caption{CARMA CO(3--2) spectrum of ACT-S\,J0210+0016, rebinned to a resolution of $40\,{\rm km\,s^{-1}}$ and plotted relative to a systemic redshift of $z = 2.55293$.  The dashed line represents the $(2.32 \pm 0.53)\,{\rm mJy}$ continuum, for
 which the line flux and luminosity have been corrected.}
 \label{Y}
\end{figure*}

\begin{figure*}
 \includegraphics[trim=0.5cm 3cm 0.5cm 6cm,width=3.5in]{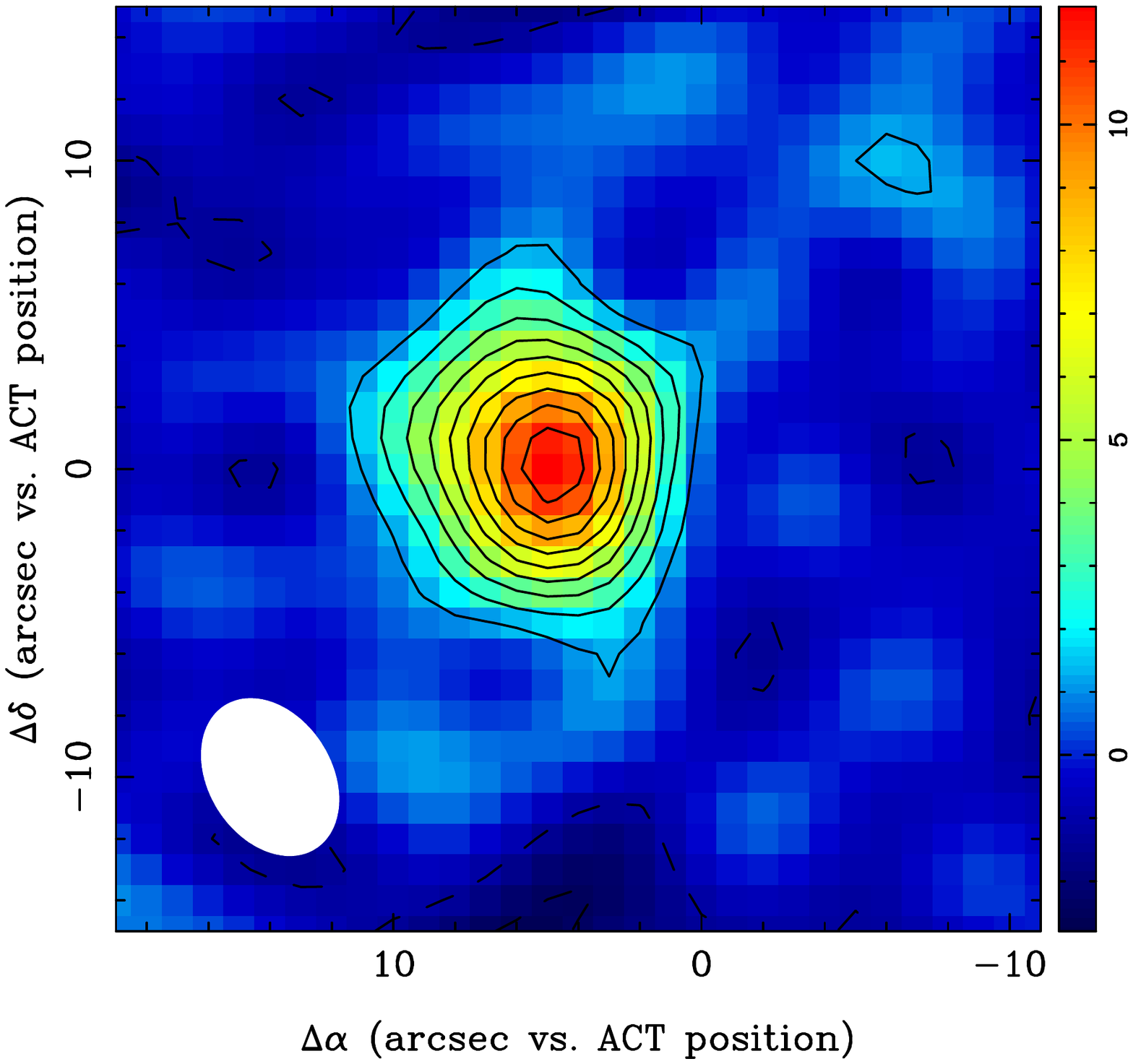} 
 \caption{CARMA CO(3--2) integrated intensity map of ACT-S\,J0210+0216, with coordinates plotted relative to the phase centre (i.e., the ACT 218\,GHz centroid position).  Contours are multiples of $1.5\sigma = 1.2\,{\rm Jy\,beam^{-1}\,km\,s^{-1}}$; negative contours are dashed. The synthesized beam is shown at lower left.}
 \label{Z}
\end{figure*}

Figures \ref{Y} and \ref{Z} show a clear detection of a spectral line source
within the uncertainty of the ACT position measurement and at the expected
frequency, confirming the redshift of the source. Spatially, a Gaussian
fit to the CO(3--2) moment map yields a centre at J2000 coordinates
$\ang{02;09;41.14} +00^h15^m57.9^s$ and a deconvolved size of $4.62^{\prime\prime}
\times 3.48^{\prime\prime}$ at a position angle of $-31.7$ degrees; these are
consistent with the parameters of the brightest radio component (A) in the
high-resolution radio continuum imaging of \citet{Geach15}. A first moment map of the CO(3--2) line shows no evidence of a significant velocity gradient. Spectrally, the
detection is centred at the same redshift as the CO(1--0) line, now resolved
into an asymmetric profile spanning approximately $450\,{\rm km\,s^{-1}}$. The zeroth
moment map shown in Figure \ref{Z} is integrated over a range of $-237$ to
$+221\,{\rm km\,s^{-1}}$ relative to line centre yielding a total line flux
of $18.2 \pm 2.0\,{\rm Jy\,km\,s^{-1}}$ and a total (apparent) line luminosity
$\mu L^\prime_{\rm CO(3-2)} = (6.5 \pm 0.7) \times 10^{11}\,{\rm
  K\,km\,s^{-1}\,pc^2}$.  Both flux and luminosity have been corrected for an
underlying continuum of $2.32 \pm 0.53\,{\rm mJy}$, estimated from the four
wideband spectral windows that symmetrically bracket the CO(3--2) line.
The ratio between the apparent CO(3--2) and CO(1--0) line luminosities is
$r_{3,1} = 1.4 \pm 0.2$, unphysically high compared to expectation for optically
thick CO lines emerging from the same volume and well above the more typical
values of 0.5--1.0 seen for star-forming galaxies and active galactic nuclei
at higher redshift \citep{Harris10,Ivison11,Riechers11,Bothwell13}. A plausible 
explanation is that the two lines are {\it not} emerging from the same volume,
due to differential lensing, in which the star-forming gas traced by the
higher-$J$ line is also more highly magnified \citep[e.g.,][]{Serjeant12}.


\section{SED fitting with other models}
\label{SEDfitting}
\begin{figure*}
  \centering
 \includegraphics[width=7.0in]{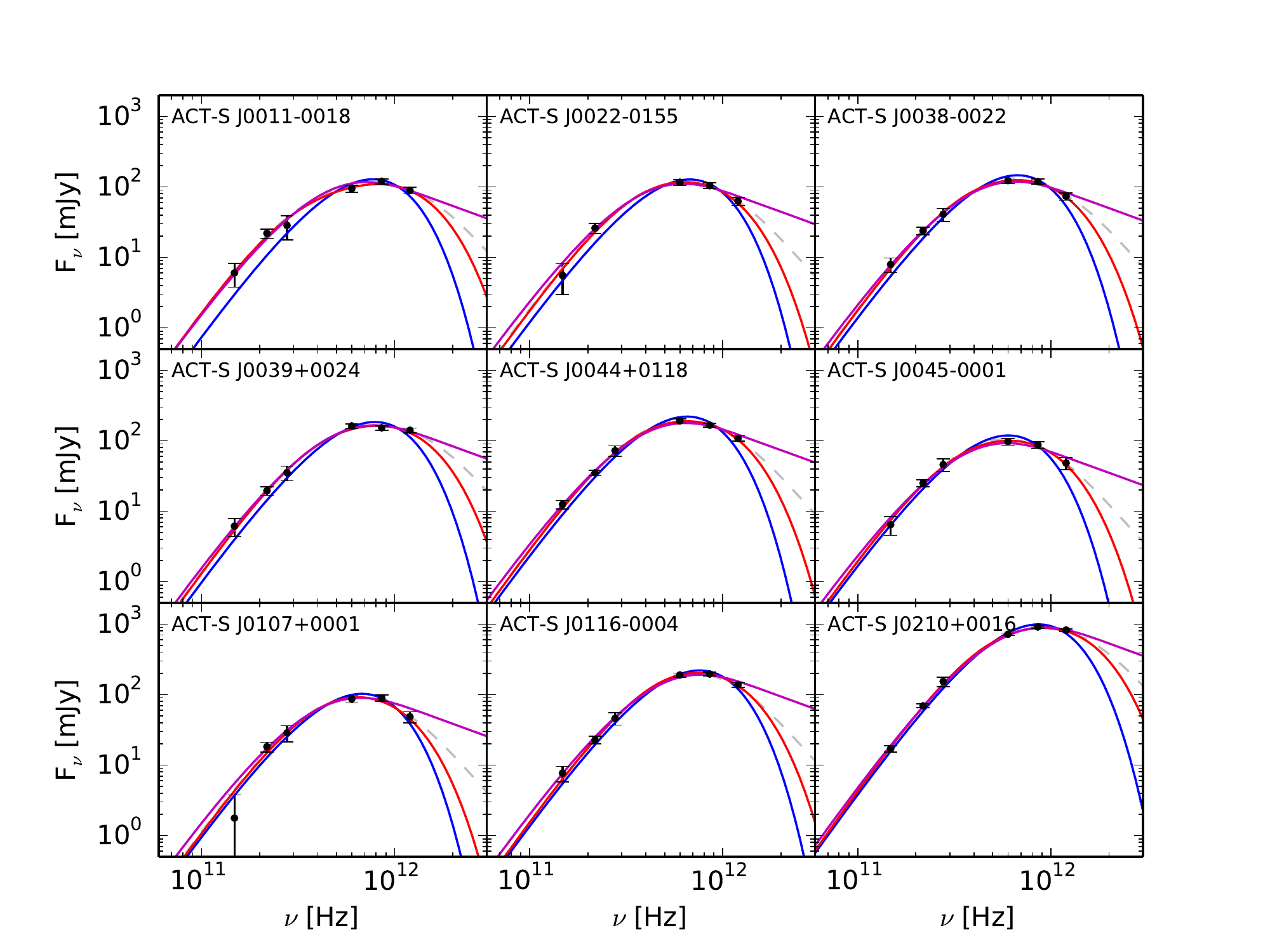} 
 \caption{SED data and best fit models for the ACT-selected lensed DSFG candidates. As plotted, the errors include systematic error (e.g., calibration uncertainties) in addition to instrument and confusion noise. Blue curves are the best-fitting results for the optically-thin, single-temperature  model. Magenta curves are the best-fitting results of the optically thin power-law temperature model. Red curves are the best-fitting results for the single-temperature model without the assumption of optical thin dust. The grey dashed curves are the best-fitting results for the fiducial  power-law temperature model without the assumption of optical thin dust (same as in Figure \ref{SEDs}).  }
\label{SEDs_appendix}
\end{figure*}

In this appendix we present the results of fits to our data using three other SED models: (1) a single-temperature model with optically thin dust, (2) a model with a power-law dust temperature distribution and optically thin dust, and (3) a single-temperature model without the assumption of optically thin dust. A discussion of the results and comparisons to the fiducial model (power-law temperature distribution, no optically thin assumption) are given in Section \ref{sec:other_models}. The numerical fit results are shown in Tables \ref{1Tthin_results}, \ref{plthin_results}, and \ref{1Tthick_results}. The models are plotted with the data in Figure \ref{SEDs_appendix}.

\begin{table*}
  \centering
  \footnotesize
  \begin{tabular}{@{}cccccccc@{}}
  \hline
  ID & $z$ &$T$ &$\log_{10}(\, \mu M_{\rm d} / {\rm M}_{\odot})$  &$ \log_{10}(\mu L_{\rm IR} / {\rm L}_{\odot})$&$\mu {\rm SFR}$ &$\chi^2 / N_{\rm dof}$\\ 
	  &	& [K]		&      & &[${\rm M}_{\odot}{\rm yr}^{-1}$]& &\\
\hline

ACT-S J0011$-$0018	&$3.6^{+0.8}_{-0.7}$		&$34.6^{+5.7}_{-5.1}$	&$9.27^{+0.19}_{-0.20}$	  &$13.34^{+0.21}_{-0.23}$ &$2300^{+1400}_{-890}$       &18.89/3 \\
ACT-S J0022$-$0155	&$4.7^{+1.0}_{-0.9}$	    &$37.1^{+5.7}_{-5.2}$	&$9.29^{+0.20}_{-0.19}$   &$13.77^{+0.19}_{-0.20}$ &$6200^{+3900}_{-2600}$       &9.35/2	\\
ACT-S J0038$-$0022	&$4.9^{+0.9}_{-0.8}$		&$38.0^{+5.8}_{-5.2}$	&$9.32^{+0.19}_{-0.20}$   &$13.86^{+0.18}_{-0.19}$ &$7700^{+3900}_{-2600}$      &14.25/3  \\
ACT-S J0039+0024	&$3.9^{+0.8}_{-0.7}$		&$37.4^{+5.9}_{-5.4}$	&$9.31\pm0.20$            &$13.81^{+0.19}_{-0.22}$ &$6800^{+3800}_{-2400}$ 	    &20.19/3  \\
ACT-S J0044+0118	&$5.8^{+1.1}_{-1.0}$		&$42.6^{+6.3}_{-6.0}$	&$9.36^{+0.21}_{-0.19}$   &$14.20^{+0.17}_{-0.19}$ &$16800^{+8000}_{-5400}$	    &25.47/3 \\
ACT-S J0045$-$0001	&$5.4^{+1.1}_{-0.9}$		&$37.4^{+6.0}_{-5.0}$	&$9.34^{+0.20}_{-0.21}$   &$13.84\pm0.19$          &$7300^{+4000}_{-2600}$      &12.30/3\\
ACT-S J0107+0001   &$4.4^{+0.9}_{-0.8}$        &$35.1^{+5.4}_{-5.0}$   &$9.27^{+0.20}_{-0.19}$   &$13.61^{+0.19}_{-0.21}$  &$4300^{+2400}_{-1500}$	        &7.23/3\\
ACT-S J0116$-$0004	&$4.4^{+0.9}_{-0.8}$	    &$40.0^{+6.3}_{-5.5}$	&$9.34\pm0.20$            &$14.00^{+0.18}_{-0.20}$ &$10600^{+5400}_{-3600}$ 	&7.74/3\\
ACT-S J0210+0016    &$5.4^{+1.0}_{-0.9}$	    &$53.2^{+8.5}_{-7.4}$   &$9.49\pm0.21$            &$14.91\pm0.18$          &$86100^{+44300}_{-29200}$  &21.0/3 \\
\hline
ACT Sample    &$4.6^{+1.2}_{-1.1}$             &$38.1^{+7.6}_{-5.9}$   &$9.32^{+0.23}_{-0.21}$   &$13.87^{+0.40}_{-0.29}$  &$7900^{+11900}_{-3800} $ &136.4/26\\
\hline
\end{tabular}
 \caption{ Results for the single-temperature, optically thin model (Equation 8).}
 \label{1Tthin_results}
\end{table*}

\begin{table*}
  \centering
  \footnotesize
  \begin{tabular}{@{}cccccccc@{}}
  \hline
  ID & $z$ &$T_{\rm c}$ &$\log_{10}(\, \mu M_{\rm d} / {\rm M}_{\odot})$  &$ \log_{10}(\mu L_{\rm IR} / {\rm L}_{\odot})$&$\mu {\rm SFR}$ &$\chi^2 / N_{\rm dof}$\\ 
	  &	& [K]		&      & &[${\rm M}_{\odot}{\rm yr}^{-1}$]& &\\
\hline

ACT-S J0011$-$0018	&$6.0^{+1.1}_{-1.0}$		&$31.5^{+4.6}_{-4.3}$	&$9.23^{+0.20}_{-0.19}$   	&$14.43^{+0.14}_{-0.16}$ &$29000^{+11000}_{-8800}$        &4.63/3 \\
ACT-S J0022$-$0155	&$7.4^{+1.3}_{-1.2}$	    &$32.7^{+4.8}_{-4.4}$	&$9.25\pm0.20$          	&$14.54^{+0.14}_{-0.15}$ &$37000^{+14000}_{-11000}$       &3.59/2	\\
ACT-S J0038$-$0022	&$6.9^{+1.3}_{-1.1}$		&$32.6^{+4.8}_{-4.5}$	&$9.25^{+0.21}_{-0.20}$     &$14.54^{+0.14}_{-0.15}$ &$37000^{+14000}_{-11000}$  &2.52/3  \\
ACT-S J0039+0024	&$5.7^{+1.0}_{-0.9}$		&$32.8^{+4.7}_{-4.5}$	&$9.25^{+0.20}_{-0.19}$ 	&$14.56^{+0.13}_{-0.16}$ &$38000^{+13000}_{-12000}$  &2.69/3  \\
ACT-S J0044+0118	&$7.9^{+1.3}_{-1.1}$		&$35.8^{+5.2}_{-4.4}$	&$9.30^{+0.19}_{-0.20}$     &$14.82\pm0.13$        &$70000^{+24000}_{-18000}$	&4.23/3 \\
ACT-S J0045$-$0001	&$7.6^{+1.3}_{-1.2}$		&$32.0^{+4.7}_{-4.2}$	&$9.24\pm0.20$		     &$14.50\pm0.14$        &$33000^{+12000}_{-9000}$      &3.35/3\\
ACT-S J0107+0001&  $6.1^{+1.2}_{-1.1}$          &$30.0^{+4.6}_{-4.1}$   &$9.24\pm0.20$           &$14.30^{+0.15}_{-0.16}$    &$21000^{+8700}_{-6500}$ &6.51/3\\
ACT-S J0116$-$0004	&$6.1^{+1.1}_{-1.0}$	   &$34.3^{+5.2}_{-4.6}$	&$9.27\pm0.20$		     &$14.68^{+0.14}_{-0.15}$ &$51000^{+19000}_{-15000}$  &4.19/3\\
ACT-S J0210+0016   &$5.9^{+1.4}_{-0.5}$	   &$39.8^{+8.2}_{-8.5}$  &$9.59^{+0.33}_{-0.27}$       &$15.37^{+0.18}_{-0.25}$  &$248500^{+127600}_{-108800}$  &2.65/3 \\
\hline
ACT Sample    &$6.8^{+1.5}_{-1.3}$          &$33.1^{+5.9}_{-4.9}$   &$9.27^{+0.23}_{-0.21}$ &$14.57^{+0.29}_{-0.22}$  &$39400^{+37400}_{-15100} $&34.36/26\\
\hline
\end{tabular}
 \caption{Results for the model with power-law temperature distribution and optically thin dust.}
 \label{plthin_results}
\end{table*}

\begin{table*}
  \centering
  \footnotesize
  \begin{tabular}{@{}cccccccccc@{}}
  \hline
  ID & $z$ &$T$ &$\log_{10}(\, \mu M_{\rm d} / {\rm M}_{\odot})$  & $\sqrt{\mu} \,d$ &$ \log_{10}(\mu L_{\rm IR} / {\rm L}_{\odot})$&$\mu {\rm SFR}$ &$\tau_{100}$&$\chi^2 / N_{\rm dof}$\\ 
	  &	& [K]		&       & &[${\rm M}_{\odot}{\rm yr}^{-1}$]& & &\\
\hline

ACT-S J0011$-$0018	&$2.9^{+0.8}_{-0.7}$		&$54.6^{+10.9}_{-9.3}$	&$9.53^{+0.20}_{-0.21}$	&$2.7^{+0.4}_{-0.3}$    &$13.44^{+0.24}_{-0.27}$ &$2900^{+2100}_{-1200}$        &$13.7^{+8.1}_{-5.1}$ &2.06/2 \\
ACT-S J0022$-$0155	&$3.7\pm0.8$	            &$49.5^{+9.3}_{-8.2}$	&$9.47^{+0.20}_{-0.21}$ &$3.8^{+0.7}_{-0.5}$ 	&$13.57^{+0.22}_{-0.24}$ &$3900^{+2600}_{-1600}$        &$5.9^{+3.5}_{-2.3}$   &1.06/2	\\
ACT-S J0038$-$0022	&$3.5^{+0.8}_{-0.7}$		&$50.8^{+9.5}_{-8.1}$	&$9.49\pm0.21$          &$3.7\pm0.5$	        &$13.58^{+0.22}_{-0.24}$ &$4000^{+2700}_{-1600}$ 		&$6.7^{+3.5}_{-2.3}$ &0.58/2  \\
ACT-S J0039+0024	&$2.7^{+0.7}_{-0.6}$		&$51.1^{+9.8}_{-8.2}$	&$9.48^{+0.19}_{-0.20}$ &$3.4\pm0.4$	        &$13.53^{+0.24}_{-0.27}$ &$3600^{+2600}_{-1500}$ 	&$7.6^{+4.0}_{-2.6}$  &4.11/2  \\
ACT-S J0044+0118	&$4.1^{+0.9}_{-0.8}$		&$56.4^{+9.9}_{-8.4}$	&$9.54^{+0.20}_{-0.21}$ &$4.4^{+0.6}_{-0.5}$    &$13.91^{+0.20}_{-0.21}$ &$8600^{+5000}_{-3200}$	&$5.1^{+2.4}_{-1.6}$ &1.48/2 \\
ACT-S J0045$-$0001	&$3.9\pm0.8$		        &$50.2^{+9.0}_{-9.0}$	&$9.48^{+0.21}_{-0.20}$ &$3.4\pm0.6$ 	        &$13.56^{+0.20}_{-0.23}$ &$3800^{+2300}_{-1400}$      &$6.6^{+4.1}_{-2.5}$ &0.84/2\\
ACT-S J0107+0001& $3.1^{+0.8}_{-0.7}$          &$43.3^{+8.6}_{-7.6}$    &$9.41^{+0.20}_{-0.21}$ &$3.7^{+1.2}_{-0.7}$ 	&$13.31^{+0.23}_{-0.26}$ &$2200^{+1500}_{-900}$	       &$5.4^{+4.0}_{-2.8}$  &4.30/2\\
ACT-S J0116$-$0004	&$3.2^{+0.8}_{-0.6}$	   &$50.0^{+8.9}_{-7.8}$	&$9.48^{+0.19}_{-0.21}$ &$4.4^{+0.6}_{-0.5}$ 	&$13.70^{+0.22}_{-0.24}$ &$5300^{+3500}_{-2100}$  &$4.4^{+2.3}_{-1.6}$ &0.69/2\\
ACT-S J0210+0016   &$2.3^{+0.9}_{-0.5}$	       &$50.5^{+13.4}_{-8.1}$   &$10.07^{+0.17}_{-0.27}$&$6.8\pm0.6$            &$14.12^{+0.34}_{-0.28}$ &$14000^{+16600}_{-7600}$  &$3.5^{+1.6}_{-1.0}$ &2.29/2 \\
\hline
ACT Sample     &$3.4^{+1.0}_{-0.8}$              &$52.1^{+12.5}_{-9.6}$   &$9.50\pm0.21$            &$3.8^{+1.3}_{-0.8}$ & $13.60^{+0.39}_{-0.32}$  &$4200^{+2600}_{-1600} $&$5.9^{+4.9}_{-2.5}$&17.3/17\\
\hline
\end{tabular}
 \caption{ Results for the single-temperature model without the assumption of optically thin dust (Equation 4). }
 \label{1Tthick_results}
\end{table*}

\end{document}